\documentclass[twocolumn]{aa}

\usepackage{txfonts}
\usepackage{graphicx}
\usepackage{natbib}
\usepackage{mathrsfs}
\bibpunct{(}{)}{;}{a}{}{,}

\begin{document}

\title{Dense core formation by fragmentation of velocity-coherent filaments in L1517
\thanks{Based on observations carried out with the FCRAO 14m and IRAM 30m telescopes.
IRAM is supported by INSU/CNRS (France), MPG (Germany) and IGN (Spain).}}

\titlerunning{Fragmentation of velocity coherent filaments in L1517}

\author{A. Hacar 
\and
M. Tafalla 
}

\institute{Observatorio Astron\'omico Nacional (IGN),
Alfonso XII 3, E-28014 Madrid,
Spain
}

\offprints{A. Hacar \email{a.hacar@oan.es}}
\date{Received 6 April 2011 / Accepted 24 June 2011}

\abstract
{Low-mass star-forming cores differ from their
surrounding molecular cloud in turbulence, shape, and density
structure.
}
{We aim to understand how dense cores form out of the less dense 
cloud material
by studying the connection between these two regimes.
}
{We observed the L1517 dark cloud in C$^{18}$O(1--0), 
N$_2$H$^+$(J=1--0), and SO(J$_{\mathrm{N}}$=3$_2$--2$_1$) with the
FCRAO 14m telescope, and in the 1.2mm dust continuum with the IRAM 30m
telescope. 
}
{Most of the gas in the cloud lies in four filaments that have
typical lengths of 0.5~pc. Five starless cores are embedded
in these filaments and have chemical compositions indicative of different evolutionary stages.
The filaments have 
radial profiles of C$^{18}$O(1--0) emission with
a central flattened region and a power-law tail, and
can be fitted approximately as isothermal,
pressure-supported cylinders.
The filaments, in addition, are
extremely quiescent. They have
subsonic internal motions and are coherent in velocity over 
their whole length. 
The large-scale motions in the filaments 
can be used to predict the velocity inside the cores,
indicating that core formation has not decoupled the 
dense gas kinematically from its parental material.
In two filaments, these large-scale motions consist of
oscillations in the velocity centroid, and
a simple kinematic model suggests
that they may be related to core-forming flows.
}
{Core formation in L1517 seems to have occurred in two steps. 
First, the subsonic, velocity-coherent filaments have condensed
out of the more turbulent ambient cloud. Then, the cores
fragmented quasi-statically and inherited the kinematics
of the filaments. Turbulence dissipation has therefore occurred
mostly
on scales on the order of 0.5~pc or larger, and seems to have 
played a small role in the formation of the individual cores.
}

\keywords{Stars: formation - ISM: clouds - molecules - kinematics and
dynamics - structure - Radio lines: ISM}

\maketitle

%

\section{Introduction}

Dense cores in
nearby dark clouds are the
birth sites of solar-mass stars,
and they represent the simplest environments where stars 
can be formed. The study of the internal structure of these cores 
offers a unique opportunity to determine
the initial conditions of low-mass star formation,
and for this reason an intense observational effort is under way
to characterize cores in detail
(see recent reviews by \citealt{DIF07,WAR07,BER07}).

Compared to their turbulent parent clouds 
dense cores are very quiescent.
Molecular line observations in tracers like ammonia reveal 
that the gas in a dense core has
subsonic internal motions, indicating that 
thermal pressure 
contributes more than turbulence in
providing support against gravity \citep{MYE83_2}.
In addition, the gas motions inside a core are ``coherent,'' 
in the sense that the observed linewidth does not depend on radius
\citep{GOO98}, and
thus deviates from the linewidth-size
relation characteristic of the large-scale gas in 
a cloud \citep{LAR81}.
Also in contrast to the irregularly shaped cloud gas, 
cores often display
a regular, close-to-spherical geometry \citep{MYE91}. 
A number of cores even present centrally concentrated
density profiles close to those predicted by
hydrostatic equilibrium models \citep{ALV01}, again suggesting that 
the core gas conditions differ significantly from those of the cloud 
gas on larger scales.

How quiescent, centrally-concentrated cores form out of 
the more turbulent, less-dense cloud material is still a 
matter of debate.
A number of core-formation mechanisms have been proposed
over the years, ranging from the quasi-static loss
of magnetic support via ambipolar diffusion \citep{SHU87,MOU99}
to the
rapid dissipation of turbulence due to supersonic shocks 
\citep{PAD01,KLE05,VAZ05}.
Observationally, it is unclear which of these
proposed scenarios can fit the variety of existing
constraints. Ambipolar diffusion models, for
example, seem to predict stronger magnetic fields
than observed \citep{CRU10}, and they
usually predict 
contraction times that significantly exceed those inferred from
observed core lifetimes and chemical clocks 
\citep{LEE99b,TAF02}.
Models of core formation by shocks,
on the other hand, predict
linewidths and velocity displacements between tracers 
that are larger than commonly observed in cores 
\citep{WAL04,KIR_H07}.

Observational progress in constraining the mechanism of core formation
requires probing the connection between
the dense cores and the less-dense material
that surrounds them. This less-dense material
likely represents the gas out of which the cores
have condensed, so by comparing its
geometry, density, and kinematics
with those of the core gas, we could infer
the physical changes involved in the
formation of a dense core. Unfortunately, studying the
cloud-to-core transition
requires combining observations of tracers
sensitive to different density regimes,
and until recently, such multi-tracer analysis has
been hindered by a number of inconsistencies
between the emission from different molecules
\citep{ZHO89,LEM96}.

Work carried out over the past decade has revealed that
most inconsistencies between tracers result from chemical
changes that occur in the gas at dense-core densities
and, in particular, from the selective freeze-out of molecules
onto cold dust grains 
\citep{KUI96,CAS99,BER02,TAF02,AIK05}.
As a result of this work, a reasonably consistent picture
of the chemical behavior of the different gas tracers
has emerged.
According to this picture, most carbon-bearing molecules
(including CO and CS)
disappear from the gas phase at densities close to a
few $10^4$~cm$^{-3}$, while nitrogen-bearing molecules
like NH$_3$ and N$_2$H$^+$ survive undepleted 
up to densities that are at least one order of magnitude higher.
In addition, several
species like SO and C$_2$S 
present significant abundance enhancements at
early evolutionary times, but end up freezing out onto
the grains at similar densities to 
the C-bearing species (see \citealt{BER07} for a review of core
chemistry).

This new understanding of molecular chemistry 
at intermediate gas densities now makes it possible
to combine observations of different tracers and to
finally explore the transition
from cloud gas to core material self-consistently. 
An excellent region for investigating this transition
is the L1517 dark cloud in the Taurus-Auriga molecular complex
(see \citealt{KEN08} for a Taurus overview).
L1517 appears in optical images as a region of 
enhanced obscuration associated with the reflection nebulosity
from the young stars 
AB Aur and SU Aur \citep{LYN62,STR76,SCH79}.
The CO observations of the cloud have shown
that the extended gas consists of several
filamentary components that extend to the northwest of the 
nebulosity and occupy a region of about $20' \times 10'$ coincident
with the optical obscuration \citep{HEY87}.
Embedded in these components lies a collection of several
dense cores that are easily distinguished
in the optical images thanks to the contrast provided
by the bright nebulosity from
AB Aur and SU Aur, which lie at least 0.3~pc in projection 
from the cores \citep{SCH79}.
Radio observations of these cores reveal the
physical properties typical of the
Taurus-Auriga core population \citep{BEN89}, and a prominent
member of the group, L1517B, has been the subject
of detailed study owing to its regular shape and clear
pattern of molecular freeze out \citep{TAF04,TAF06}.

A notable feature of the L1517 cores is that 
they all appear to be starless 
\citep{STR76,BEI86,KIR07}.
This lack of embedded protostars is probably
responsible for the L1517 cloud presenting some of the
narrowest line profiles in CO and dense gas tracers 
seen towards the Taurus-Auriga region,
and indicates that, although 
the bright PMS star AB Aur ($\sim 50$~L$_\odot$, \citealt{VAN98})
is physically related to
the cloud \citep{NAC79,DUV86},
the bulk of the L1517 material and its
embedded cores remain unperturbed by the 
energetic stellar output
\citep{LADD91}. This combination of 
quiescent state, compact size, and multiple
starless cores 
makes the L1517 cloud an ideal laboratory for studying
the process of core formation. In this paper, we present
a study of the physical conditions and
kinematics in both the dense cores and the less-dense
surrounding material using a variety of molecules
known to trace different density
regimes and chemical evolutionary stages of the cloud gas.
As will be seen, our analysis suggests that the cores
in L1517 have
formed by the gravitational contraction of
subsonic, velocity coherent gas in 0.5~pc-long filaments.

\section{Observations}

We observed the L1517 dark cloud with the FCRAO 14m radio telescope
during several sessions between December 2003 and November 2005.
We used  the 32-pixel SEQUOIA array to cover the cloud
with five submaps, each of them of 
$10'\times 10'$, and observed in on-the-fly mode.
The large
bandpass of SEQUOIA allowed observing two different
transitions simultaneously, and two passes were made
to cover the cloud first
in N$_2$H$^+$(J=1--0) and SO(J$_{\mathrm{N}}$=3$_2$--2$_1$)
and then in C$^{18}$O(J=1--0) and C$^{17}$O(J=1--0).
All observations were made in position-switching mode
using a reference position offset by 
($18'$, $-13'$) from our map center (at
$\alpha(J2000)=4^h55^m18\fs8,$ $\delta(J2000)=+30^\circ38'04''$),
and known to have negligible
C$^{18}$O(J=1--0) emission from previous frequency-switched
observations (average $\sigma(T_{mb})<0.1$~K over the SEQUOIA
footprint). The spectrometer was the DCC autocorrelator configured
to provide 1024 spectral channels of 25~kHz spacing, or
approximately 0.07~km~s$^{-1}$ at the observing frequencies.
The telescope FWHM beam size varied with frequency between
approximately $56''$ at the lowest (N$_2$H$^+$)
frequency and $47''$ at the highest (C$^{17}$O)
frequency.

During the observations, 
calibration was achieved by measuring the emission from the sky
and an ambient load every $\approx 10$ minutes, and the
derived intensity
was converted into the main beam brightness 
temperature scale
using facility-provided
main beam efficiencies close to 0.5.
The telescope pointing was checked and corrected
approximately every three hours by making five-point maps
on the SiO masers of IK Tau and Orion-IRc2. Typical
pointing errors were within $5''$ rms.

The off-line data reduction consisted in the creation
of Nyquist-sampled maps
with the {\tt otftool} program and a 
conversion 
to the CLASS format for further
analysis with the
GILDAS software ({\tt http://www.iram.fr/IRAMFR/GILDAS}).
This analysis included 
a second-degree baseline subtraction and a
spatial convolution with a Gaussian
to eliminate residual noise.
The final resolution of all the FCRAO data is $60''$,
taking the off-line convolution into account.

Because the on-the-fly FCRAO maps are relatively shallow because of the 
large area covered, we complemented the N$_2$H$^+$(1--0) 
FCRAO observations with 
a small amount of data observed with the IRAM 30m
radio telescope. 
These data were taken in frequency-switching
mode and have a higher angular resolution than 
the FCRAO data (about $26''$), but apart from that,
they are consistent with the
FCRAO observations in both intensity and frequency calibration.
A full report of these IRAM 30m data will be
presented elsewhere as
part of an extended study of core chemistry
that includes a number of regions in addition to L1517
(Tafalla et al. in preparation). 
In the present paper, the IRAM 30m data are only used 
to complement the FCRAO observations in the velocity
analysis, which requires a high S/N threshold. 

Given the narrow lines and
small velocity variations measured towards
L1517, accurate rest-line frequencies were required
for analyzing of the spectra. 
For this reason, we used the most recent laboratory estimates
for C$^{18}$O(J=1--0) (109782.176 MHz, see
\citealt{CAZ03}) and for C$^{17}$O(J=1--0) 
(112358.990 MHz for the brightest,
J F=1 7/2--0 5/2 component, see \citealt{CAZ02}),
which have  1 kHz or better accuracy.
For N$_2$H$^+$ and SO,
no accurate laboratory measurements are available,
so we used astronomical estimates. 
Following \citet{PAG09}, we assumed a
frequency of
93173.764 MHz for N$_2$H$^+$(J$F_1$F = 123--012), with
an estimated uncertainty of 4~kHz,
and following \citet{TAF06}, we assumed a 
frequency of 99299.890~MHz for 
SO(J$_{\mathrm{N}}$=3$_2$--2$_1$), with an
estimated uncertainty of 10~kHz.

\begin{figure*}
\centering
\resizebox{12cm}{!}{\includegraphics{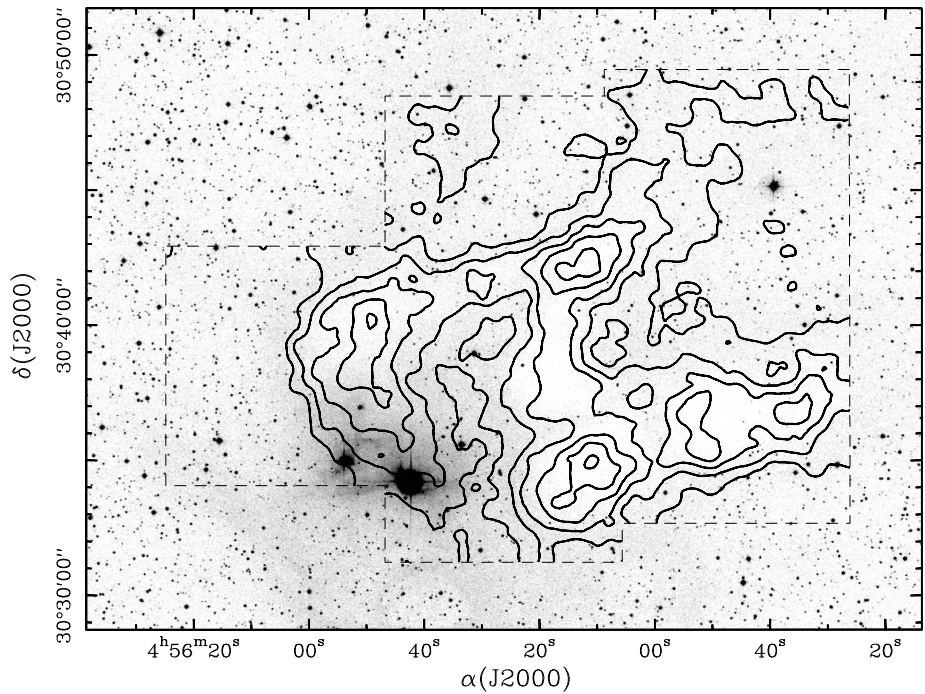}}
\caption{Map of the C$^{18}$O(1--0) integrated intensity
towards the L1517 cloud superposed to the red DSS image.
The emission has been integrated over the whole velocity range
of detection (V$_{LSR}$ = 5.25-6.45~km~s$^{-1}$).
First contour and contour interval are 0.22~K~km~s$^{-1}$.
The two bright stars surrounded by nebulosity and 
located near the southeastern edge of the C$^{18}$O(1--0)
emission are AB Aur (brightest) and SU Aur.
\label{DSS_C18O}}
\end{figure*}

Additional observations of the dense cores in L1517
were carried out in the 1.2mm continuum using the MAMBO array
on the IRAM 30m radio telescope during several sessions
between December 1999 and January 2005. The data for
core L1517B have already been presented  in \citet{TAF04},
while the data for cores L1517A and L1517C are newly reported here.
(L1517D was not observed.) 
In all cases, the observations were done in 
on-the-fly mode scanning the telescope 
in azimuth at a speed of $4''$s$^{-1}$, and using wobbler
switching with a
period of 0.5s and a throw of $53''$ or $70''$.
The raw data were corrected for atmospheric attenuation using sky dips 
usually made before and/or after the source observation,
and the absolute intensity calibration was achieved using
facility-provided factors derived from observations of planets. 
The older L1517B data were reduced with the NIC software
using no method of noise reduction (to avoid filtering the
extended emission), while the newer L1517A and L1517C data were 
reduced with MOPSIC using a noise reduction method
optimized to recover the extended emission.
The intrinsic 
beam size of all the continuum data is $11''$, although 
the maps were later convolved to an equivalent resolution of $20''$
to eliminate high spatial-frequency noise.

\section{C$^{18}$O data: the extended cloud}

Figure \ref{DSS_C18O} presents our map of C$^{18}$O(1--0)
emission integrated over the full velocity range of
detection superposed to the red DSS image. 
This map is in good agreement with the previous
map by \citet{HEY87}, and it shows how the 
C$^{18}$O(1--0) emission extends to the 
northwest of AB and SU Aur, the brightest stars in the DSS image.
These PMS stars 
are physically associated with the L1517 cloud, as 
indicated by the heating of the gas in their vicinity
\citep{NAC79,DUV86}, but their effect must be
highly localized since the data described in the next sections
show no evidence of any kinematic interaction between
the gas and the stars. 
This lack of kinematical interaction, together with the
fact that the northwest elongation of the L1517 cloud
continues in the maps of 
\citet{DUV86} to scales as large as 4pc (or more than 5
times the size of our map) and involves additional
dark clouds such as L1496, L1505, L1513,
and L1515, suggests that the distribution of gas
in L1517 is intrinsic to the cloud, and that it has not
been sculpted by the PMS stars. In this sense, the
L1517 cloud seems a quiescent remnant of a larger 
gas cloud that gave rise to AB Aur, SU Aur, and
an additional group of PMS stars located in their
vicinity \citep{LUH09}.

\subsection{Filament identification}\label{fils}

\begin{figure*}
\centering
\resizebox{18cm}{!}{\includegraphics[angle=270]{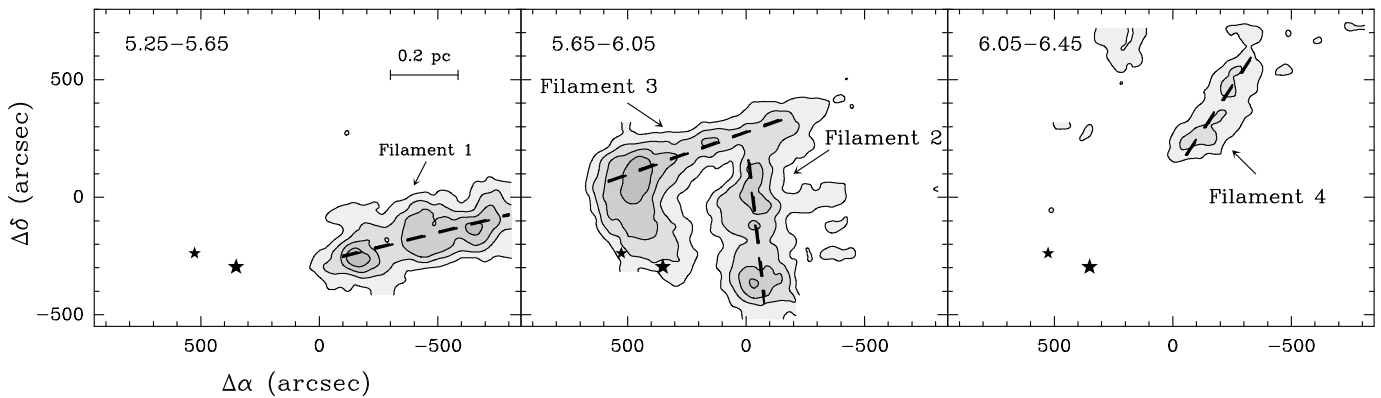}}
\caption{Maps of C$^{18}$O(1--0) intensity integrated every 0.4 km s$^{-1}$
to illustrate the separation of the gas into filaments. 
The dashed lines indicate our best guess of each filament axis,
and the star symbols mark the positions of AB Aur (larger) and 
SU Aur (smaller). First contour and interval are 0.2~K~km~s$^{-1}$.
Offsets are referred to $\alpha(J2000)=4^h55^m18\fs8,$ 
$\delta(J2000)=+30^\circ38'04''$. The LSR velocity range of
integration is indicated in the upper left corner of each panel.
\label{C18O_channelmap}}
\end{figure*}

A more detailed view of the gas distribution 
in L1517 comes from 
the analysis of its velocity structure.
Figure~\ref{C18O_channelmap} shows maps of 
C$^{18}$O(1--0) emission integrated every 0.4 km s$^{-1}$,
or about twice the sound speed for gas at 10K.
These maps, and others made using different velocity ranges,
show that the L1517 cloud is structured in at least four 
elongated components that are referred to 
as filaments 1 to 4.
These four filaments can also be seen in maps of extinction 
(Jouni Kainulainen, private communication), $^{13}$CO \citep{HEY87},
and SO emission (Sect.~\ref{cores}), so they must
reflect the true distribution of gas in the L1517 cloud, 
and are not mere artifacts of the 
C$^{18}$O chemistry or excitation. 

As Fig.~\ref{C18O_channelmap} shows, each filament appears in only one 
0.4~km~s$^{-1}$-wide channel map, indicating that its gas
is highly confined in velocity space. This low level of 
velocity structure results from a very quiescent state
of the gas in the filaments, both in terms of velocity dispersion and
internal velocity gradients
(to be discussed below),
and makes it possible to associate each element of emission
in the cloud
to one of the four filaments. To carry out this assignment, 
we have taken both the 
spatial location of the emission and its velocity into account and assumed that the velocity limits of the filaments
are those used in the maps of 
Fig.~\ref{C18O_channelmap} (see also Table~\ref{filaments}).
This assigning procedure works well over most of the mapped region,
but it provides ambiguous answers in 
the vicinity of offsets ($-100''$, $-300''$).
In this region, filaments 1 and 2 overlap spatially
and converge in velocity to a value close to
5.65~km~s$^{-1}$, which is our assumed boundary between
the two filaments. 
A detailed inspection of the FCRAO spectra in this region
(supplemented with additional data from the IRAM 30m telescope)
reveals that the emission from the two filaments 
becomes almost indistinguishable, either because the
filaments merge physically
or because they superpose along the line of sight 
with similar kinematic properties.
As a result,
assigning  emission in this region to either filament 1 or 
2 becomes more uncertain than in other 
places of the cloud. 
Although this may lead to some confusion in
a localized region, 
it is unlikely to affect our global
analysis of the gas kinematics, since the 
properties of the two filaments
become so similar in
terms of velocity centroid and linewidth 
that an erroneous assignment of a gas parcel 
to either filament will only add a small amount
of mass, but will not change its velocity field.

Once we have decomposed the L1517 cloud into its constituent filaments,
we use the C$^{18}$O (1--0) emission to estimate their
basic physical parameters, and we summarize the results in 
Table~\ref{filaments}.
From the maps of Fig~\ref{C18O_channelmap}, 
we estimate that 
the filaments have lengths of
550-1000 arcsec, which correspond to physical sizes of
$\sim$ 0.35-0.70 pc for our adopted distance of 144 pc (based on the
Hipparcos distance to AB Aur, \citealt{VAN98}).
These sizes represent a significant fraction of the 
total cloud length, and this reinforces the idea that the
cloud is structured as a network of filaments. To estimate
the masses of the filaments, we use the 
C$^{18}$O (1--0) emission and assume that it is optically thin
and in LTE at 10~K. For a C$^{18}$O abundance of
$1.7 \times 10^{-7}$ \citep{FRE82}, the typical
filament masses are 5-11~M$_\odot$,
which imply linear mass densities in
the range 12-17~M$_\odot$~pc$^{-1}$.

\begin{table}
\caption[]{Filaments in L1517.
\label{filaments}}
\begin{tabular}{ccccc}
\hline
\hline
\noalign{\smallskip}
 Filament  & V$_{LSR}$ range & Length$^{(1)}$ & Mass$^{(2)}$ & Cores \\
      &  \mbox{(km s$^{-1}$)}  & (pc) &  (M$_\odot$)    &  \\
\noalign{\smallskip}
\hline
\noalign{\smallskip}
1  & $[5.25-5.65]$  & 0.52 & 8.0 & A2, C \\
2  & $[5.65-6.05]$  & 0.42 & 7.2 & A1, B \\
3  & $[5.65-6.05]$  & 0.70 & 11.3 & D \\
4  & $[6.05-6.45]$  & 0.38 & 4.8 & --- \\

\hline
\end{tabular}
\begin{list}{}{}
\item[Notes:] (1) Uncorrected for projection effects; (2)
From C$^{18}$O(1--0) emission.
\end{list}
\end{table}

\subsection{Density structure of the filaments}\label{fils_density}

\begin{figure}
\centering
\includegraphics[width=7.5cm]{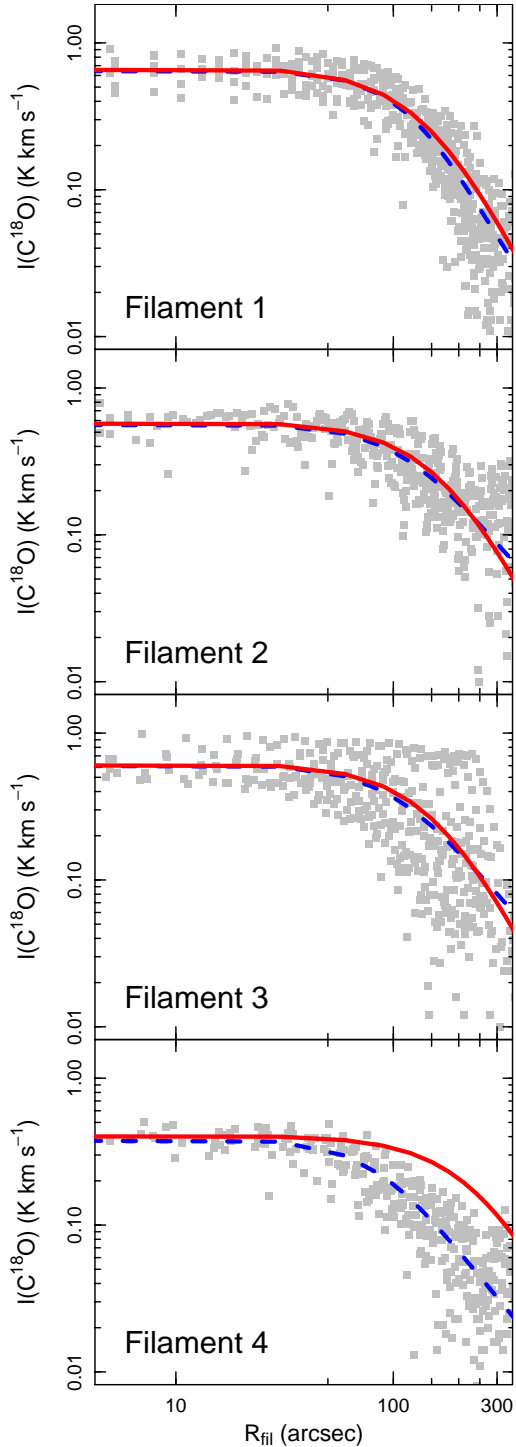}
\caption{C$^{18}$O(1--0) 
emission profiles across each of the L1517 filaments. 
The points
represent the integrated emission within the velocity
range of each filament (see Table \ref{filaments}), and R$_{fil}$
is the distance to the filament axis.
For each filament, the red solid curve
indicates the isothermal cylinder model that fits the emission towards the
axis, and the blue dashed line is the best-fit softened power law.
\label{Fils_RP}}
\end{figure}

As Fig.~\ref{C18O_channelmap} shows, the C$^{18}$O emission 
from the filaments displays a significant degree of central brightening.
This suggests that the underlying density structure of the filaments
is centrally concentrated and that the C$^{18}$O emission can
be used to determine each filament density profile.
To carry out this determination, we first defined a
central axis for each filament assuming that they are rectilinear
and show the results in Fig.~\ref{C18O_channelmap}.
As can be seen,
the rectilinear approximation is reasonable
for all filaments but number 3, which 
bends towards the location
of the PMS stars near its southeast end. Whether this
bend is related to core L1517D, with which it coincides
in position (Sect.~4), or is associated to the PMS stars
is unclear from our data. In either case, for simplicity
in the modeling,
we approximate all filaments by straight lines, with
the caveat that this approach is only a first-order
approximation.

Using the above linear axes, we 
created a profile of C$^{18}$O (1--0) 
intensity for each filament as a function of cylindrical radius. The 
result, presented in Fig.~\ref{Fils_RP},
shows that the four filaments in the cloud have
centrally concentrated distributions with
a flattened region close to the axis and an approximate
power-law tail at large distances.
Filament~3 presents the highest
dispersion of the sample due to a combination of contamination 
from filament 2 emission near its northern edge and the additional
contribution from the ``bend'' region previously discussed.
Nevertheless, given the rather irregular distribution of emission
seen in Fig.~\ref{C18O_channelmap}, it is
remarkable that the radial profile of each filament presents
relatively little scatter and that all the 
filaments seem to follow a very similar type of 
radial profile.

The regular behavior of the radial profiles
in Fig.~\ref{Fils_RP}
suggests that the underlying density structure of the filaments
can be described with a simple density law. 
To derive such a  law, we model the
C$^{18}$O (1--0) emission assuming 
that each filament has cylindrical
symmetry and that the C$^{18}$O emission arises
from gas in LTE 
thermalized at 10~K, as suggested by the study of the
L1517B core and its surrounding
envelope in \citet{TAF04}.
Under these conditions, if we assume a
density profile, together with a constant
C$^{18}$O abundance (1.7$\times$10$^{-7}$, \citealt{FRE82})
and a C$^{18}$O linewidth of 0.3~km~s$^{-1}$
(Sect.~\ref{linewidths1}),  we can
predict a radial profile of emission
that can be compared, after appropriate
beam convolution, with the observed radial profiles
shown in Fig.~\ref{Fils_RP}.

As a first guess for the filament density profiles, we use
the family of isothermal cylinders in 
pressure equilibrium with their self gravity.
This family was  originally described by 
\citet{STO63} and \citet{OST64}, and
has the simple analytic form 
\begin{equation}
\label{iso_cyl}
n(r)=\frac{n_\circ}{(1+(r/H)^2)\,^2},
\end{equation}
where $r$ is the cylindrical radius, $n_\circ$ is the central density,
\begin{equation}
\label{h_diam}
H^2=\frac{2c_s^2}{\pi G \mu n_\circ},
\end{equation}
$c_s$ is the isothermal sound speed, 
$G$ the gravitational constant, and $\mu$ the mean molecular mass.
For the case of L1517, the gas kinetic
temperature is known to be approximately
10~K \citep{TAF04}
so any isothermal cylinder 
is described by just one parameter, its central density $n_\circ$.
This means that once we have set the central density to
a value that fits the intensity at zero radius, we do not
have any free parameter left to control the rest of the radial
profile, and in particular, the width at half maximum 
of the emission is automatically determined by the $H$ value.

There are a number of motivations for using the isothermal cylinder family as
a first choice to fit the C$^{18}$O radial profiles.
In addition to the
inherent simplicity and assumption of equilibrium of the model, the isothermal
cylinder is attractive for presenting both a central density flattening
and a power-law tail, which are two characteristics of
the observed radial profiles. The isothermal cylinder, in
addition, has a mass per unit length
that is independent of the central density and is only a function
of the gas temperature \citep{OST64}. For the 10 K assumed for the
gas in L1517, this mass per unit length is approximately
16~M$_\odot$~pc$^{-1}$, which is reasonably close to the mass
per unit length of the filaments  estimated in 
Table~\ref{filaments}.

The results of our isothermal cylinder fits to the filament profiles in
L1517 are shown in Fig.~\ref{Fils_RP}.
These fits were selected to match the 
C$^{18}$O emission towards the filament axis,
so the radius of half maximum emission
is a direct prediction from the model
so is not controlled by our fit.
As can be seen, the 
predicted filament widths 
are in reasonable agreement with the observations,
with only filament 4 being
clearly narrower than predicted by the model.
In addition, the 
central densities predicted by the fits range
from 3$\times$10$^{3}$ to
1$\times$10$^{4}$ cm$^{-3}$ (see 
Table~\ref{Fils_fits}), and these values agree with
previous estimates of the gas density in the extended
part of the L1517 cloud \citep{TAF04}.

To check the  consistency of the parameters derived with the
isothermal cylinder model and to provide a better match
to the radial profiles of the filaments,
we have fitted the C$^{18}$O emission
with an alternative family of profiles that also have 
central flattening and asymptotic power-law behavior.
This family consists of softened power laws described by 
\begin{equation}
\label{threepar_cyl}
n(r)=\frac{n_\circ}{1+(r/r_{1/2})\,^\alpha},
\end{equation}
where $n_\circ$ is again the central density, $r_{1/2}$ the half-density
radius, and $\alpha$ the asymptotic power index. This family of profiles 
has been shown to fit
the density structure of starless cores, which also present
flattened central
regions and asymptotic power-law behavior \citep{TAF04}.
In contrast to the equilibrium cylinder, the softened power law
has three
free parameters, so it fits  the central density
and the width of the filament independently.
By comparing the results from the softened power law
fits with those from the isothermal cylinders,
we can now test how self consistent the one-parameter 
isothermal-cylinder fits are.

Figure~\ref{Fils_RP} shows the softened power law fits
derived with a chi-squared minimization algorithm. As 
can be seen, these fits are almost
indistinguishable from the isothermal cylinder fits 
for filaments 1, 2, and 3,
and the only clear difference 
between the two families of fits
occurs in filament 4. 
A more quantitative comparison
between the fits 
comes from examining the derived
central density and filament width
in the two families.
Table~\ref{Fils_fits} shows that these parameters 
differ by less than 10 \% on average for the first 
three filaments, indicating again that 
the isothermal cylinder and softened power-law
fits are
equivalent within the scatter of the observations.
In filament 4, the softened power-law fit is clearly
superior to the (poor) isothermal cylinder fit,
and the difference between the derived parameters
in the density law is approximately a factor of 2.

\subsection{Implications and limitations of the modeling}

Our analysis of the L1517 filaments
adds to a number of previous studies of filamentary
structures.
With few exceptions, like that of 
\citet{JOH03} who fitted the 850~$\mu$m
emission from the infrared-dark cloud G11.11-0.12,
most previous work has found
significant deviations between the
radial profiles of the filaments and the
prediction from the isothermal cylinder model, usually
because the observed asymptotic
power-law behavior is flatter than the
$r^{-4}$ predicted by the isothermal cylinder
model \citep{ALV98,STE03,AND10,ARZ11}.
For the filaments in L1517, our
radial profiles do not extend  
far enough in radius to sample the $r^{-4}$ asymptotic
behavior, as can be seen in Fig.~\ref{Fils_RP} from
the isothermal
cylinder models being practically indistinguishable from softened
power-law profiles with an asymptotic behavior close
to $r^{-2.7}$ (Table \ref{Fils_fits}). As a result, 
testing whether the L1517
filaments follow the prediction from the model or
deviate from it cannot be done with our data, and
it requires extending the radial profiles
to larger radii, preferably using more robust techniques, such as extinction
measurements or dust continuum observations.

Even if the L1517 filaments do not follow the expected 
asymptotic behavior, it is striking that they approximately
fit the expectations from the isothermal cylinder model both in
width and mass per unit length. Clearly there are a number
of limitations in our study, like the use of the freeze-out prone 
C$^{18}$O molecule as a column density tracer and our ignoring 
of projection effects, so there is probably room for a factor of 2 
uncertainty in the results. Also, as we will see
in section~\ref{coreformation}, the size scale of fragmentation suggests
that there may be a contribution from additional forces,
like external pressure or magnetic fields (which could even
flatten the radial profile, e.g., \citealt{FIE00a}).
Still, the very quiescent state of the cloud described
in the following sections indicates that the gas
in L1517 cannot be too far from a state of equilibrium, 
as otherwise it would quickly develop supersonic motions,
(e.g., \citealt{BUR04}) and would not fragment into
well-separated cores \citep{INU97}.
As pressure forces dominate,
it  seems reasonable to expect that the distribution of
mass in the filaments bears some similarity to the
prediction from the isothermal cylinder model.

\begin{table}
\caption[]{Fits to the C$^{18}$O radial profiles.$^{(1)}$
\label{Fils_fits}}
\begin{tabular}{c|cc|ccc}
\hline
\hline
\noalign{\smallskip}
 & \multicolumn{2}{|c|}{Isothermal cylinder} &
\multicolumn{3}{c}{Softened power law} \\
\noalign{\smallskip}
\hline
\noalign{\smallskip}
 Filament  & $n_\circ$ & r$_{1/2}$$^{(2)}$ &  $n_\circ$ & r$_{1/2}$ & 
 $\alpha$ \\
  &  \mbox{(cm$^{-3}$)}  & (arcsec) & \mbox{(cm$^{-3}$)}  & (arcsec) &  \\
\noalign{\smallskip}
\hline
\noalign{\smallskip}
1  & $1\times10^4$ & 89 &$8.5\times10^3$ & 105 & 3.4 \\
2  & $7\times10^3$ & 106 &  $7\times10^3$ & 100 & 2.7 \\
3  & $8\times10^3$ & 99 & $8.5\times10^3$ & 90 & 2.7 \\
4  & $3\times10^3$ & 162 & $6.5\times10^3$ & 75 & 2.8 \\
\hline
\end{tabular}
\begin{list}{}{}
\item[Notes:] (1) Not corrected for inclination; 
(2) r$_{1/2} = (2^{1/2} - 1)^{1/2} H$
\end{list}
\end{table}

\section{Dense core population}\label{cores}

\begin{figure*}
\centering
\includegraphics{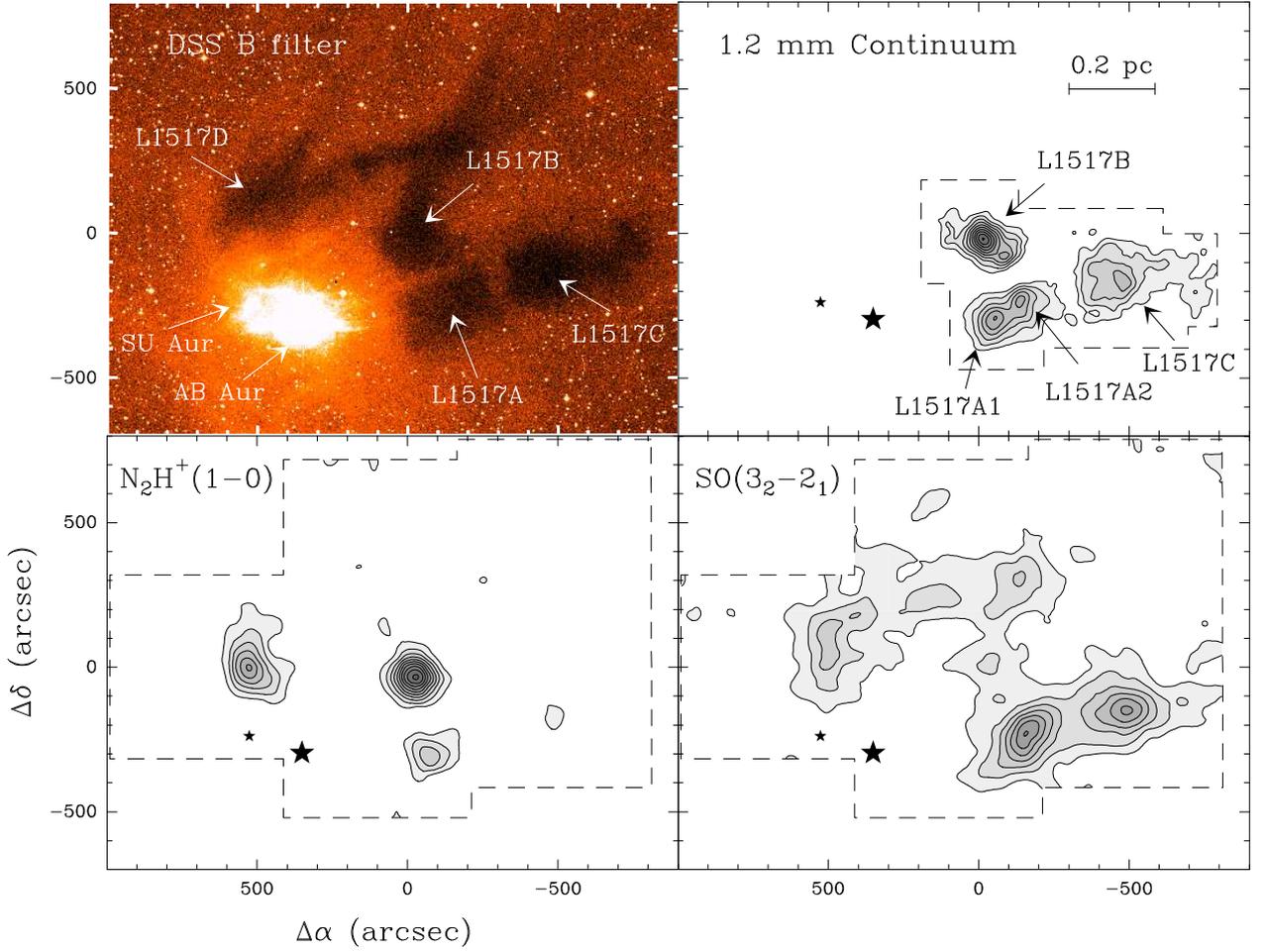}
\caption{Core population of L1517. From left to right and from
top to bottom: contrast-enhanced DSS blue image identifying the core
positions,  1.2mm
dust continuum emission, N$_2$H$^+$(1--0) integrated intensity,
and SO(3$_2$--2$_1$) integrated intensity.
Offsets and star symbols are as in Fig.~\ref{C18O_channelmap}.
First contour and spacing are 2~mJy/11$''$-beam for
1.2mm continuum, 0.16~K~km~s$^{-1}$ for N$_2$H$^+$(1--0),
and 0.08~K~km~s$^{-1}$ for SO(3$_2$--2$_1$).
To enhance the sensitivity to extended emission, the 
1.2mm continuum map has been convolved to an equivalent resolution
of $30''$, and the N$_2$H$^+$ and SO maps 
to a resolution of $75''$.
\label{densegas}}
\end{figure*}

The population of dense cores in the L1517 cloud was 
first described by \citet{SCH79}, who
identified four regions of enhanced 
obscuration in the
Palomar optical plates and named them A to D in order
of increasing right ascension (RA).
For unknown reasons, 
this original labeling scheme was later replaced
by one in which the cores
are named C, A, B, and D in order of increasing RA,
and this notation has been used by most core studies
in the past (e.g., \citealt{BEN89}). 
For consistency with previous work, we
also use this disordered core labeling  here.

Figure~\ref{densegas} presents different views of the core population 
in L1517. The optical DSS image on the top left has had its contrast 
enhanced to better show the cores as dark patches against both the
galactic stellar background and the bright diffuse emission from
AB Aur and SU Aur, which are probably slightly behind the cores.
The other panels show maps of the same region
in the 1.2mm dust continuum, N$_2$H$^+$(1--0), and
SO(3$_2$--2$_1$), three tracers that highlight different properties of the
dense, high column density gas. 
The 1.2mm continuum emission is
mostly sensitive to the column density of the cores 
(the extended emission from the filament is filtered out by the bolometer),
and its maps
provide high angular resolution views of cores A, B, and C
(core D was not mapped).
The brightest 1.2mm emission corresponds to core B, whose 
structure and chemical composition has been studied in detail by 
\citet{TAF04,TAF06}. Cores A and C have received less
attention due to their weaker emission in both high-density
tracers and mm-continuum 
\citep{BEN89,LADD91,KIR05},
so their structure is less well known.
Our mm-continuum map shows that core A is double-peaked,
while core C is single-peaked but relatively  more
extended and significantly weaker than cores A and B.
Overall, the mm-continuum map 
is characterized by a lack of
point-like components, which together with the absence of
Spitzer $24~\mu$m point sources 
\citep{KIR07} is a strong indication
that all dense cores in the cloud are starless.

The N$_2$H$^+$ and SO maps in Fig.~\ref{densegas} present 
complementary views of 
the L1517 cores. N$_2$H$^+$ is a so-called late-time molecule,
and its abundance is further enhanced when CO freezes out
(e.g., \citealt{BER07}),
so bright N$_2$H$^+$ emission is commonly associated with
evolved cores \citep{CRA05}.
As Fig. \ref{densegas} shows, the N$_2$H$^+$ emission 
is brightest towards
cores B and D, and still noticeable towards core A1.
These cores therefore seem 
more chemically evolved than cores
C and A2, which are barely detected in N$_2$H$^+$(1--0).
Such interpretation of the N$_2$H$^+$ maps
in terms of chemical evolution
is supported by the distribution of SO emission.
Observationally,
SO is known to be highly sensitive 
to molecular freeze out \citep{TAF06},
and in addition, it is predicted to decrease in abundance with
time \citep{BER97, AIK05}.
The SO emission, therefore, is expected to be anticorrelated
with that of N$_2$H$^+$, and indeed, the maps of L1517
show such behavior: the SO emission is dominated by cores
C and A2, which are weak in  N$_2$H$^+$(1--0),
while the N$_2$H$^+$-bright cores A1, B, and D
are barely distinguishable from
the SO emission of the extended cloud.

\begin{table}
\caption[]{Kinematic properties of the L1517 cores.
\label{cores_prop}}
\begin{center}
\begin{tabular}{lcccc}
\hline
\hline
\noalign{\smallskip}
 Core  & $\Delta\alpha$, $\Delta\delta$ &
V$_{lsr}(\mathrm{N}_2\mathrm{H}^+)$ &
$\Delta\mathrm{V}(\mathrm{N}_2\mathrm{H}^+)$ & $\langle|\nabla
V_{lsr}(\mathrm{N}_2\mathrm{H}^+)|\rangle$  \\
      &  ($''$, $''$)  & (km s$^{-1}$) & (km s$^{-1}$)& (km
s$^{-1}$ pc$^{-1}$)   \\
\noalign{\smallskip}
\hline
\noalign{\smallskip}
A1  & -60, -300 & 5.70 & 0.18 & 0.7$\pm$0.1 \\ 
A2$^{(1)}$  & -150, -240 & 5.57 & 0.20 & ---$^{(2)}$ \\ 
B  & -30, -30 & 5.79 & 0.22 & 0.7$\pm$0.3 \\ 
C$^{(1)}$  & -480, -150 & 5.48 & 0.18 & ---$^{(2)}$ \\ 
D  & 540, 0 & 5.88 & 0.29 & 0.9$\pm$0.4 \\ 
\hline
\end{tabular}
\end{center}
Notes: $^{(1)}$ Kinematics data from IRAM 30m observations 
(rest from FCRAO 14m).
$^{(2)}$ Not enough data to estimate gradient.
\end{table}

\subsection{Core emission modeling}\label{core_model}

\begin{figure*}
\centering
\resizebox{17cm}{!}{\includegraphics[angle=270]{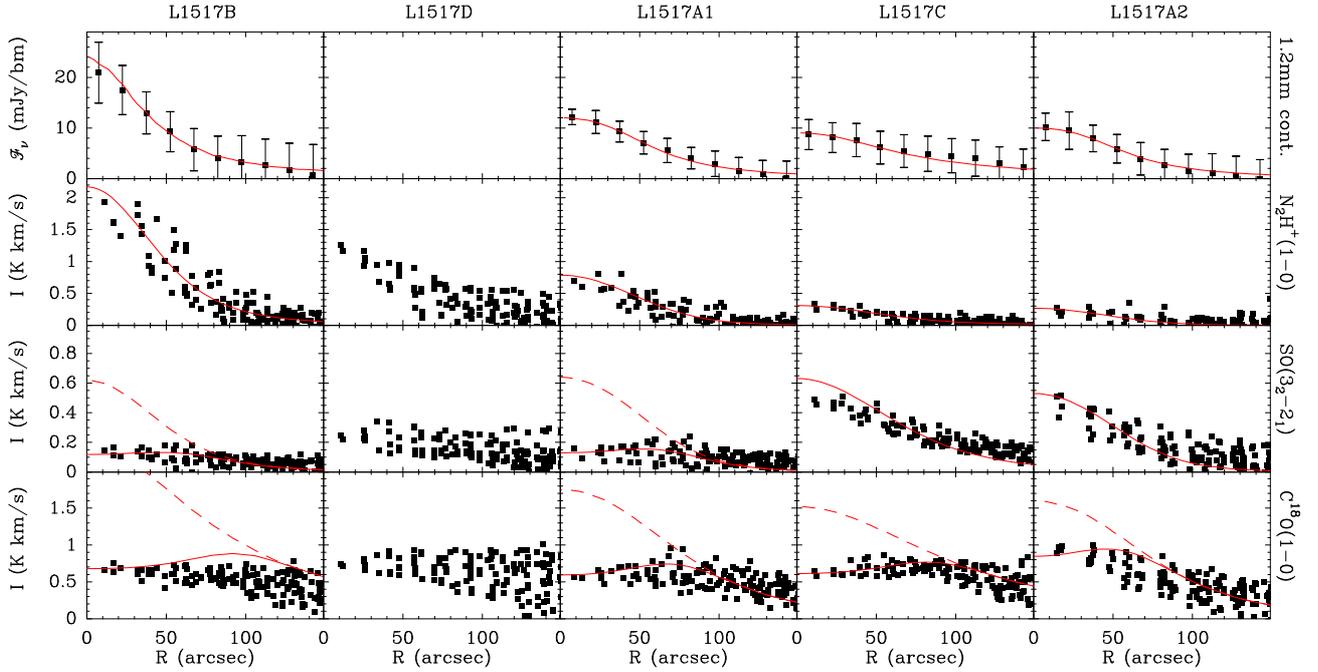}}
\caption {Radial profiles of 1.2mm continuum, N$_2$H$^+$(1--0),
SO($3_2$--$2_1$), and C$^{18}$O(1--0) emission towards the dense 
cores in L1517. 
The cores were ordered by decreasing central emission
of N$_2$H$^+$(1--0) to illustrate a possible evolutionary
sequence (see text).
The 1.2mm data were averaged over
$15''$-wide intervals to increase legibility.
In each panel, the solid red line
represents the best fit-model, and when
an additional
dashed line is presented,
the best fit required a central abundance hole
and the dashed line represents the constant abundance model
for comparison. No modeling of core D 
was attempted
for lack of mm-continuum data.
\label{cores_radial}}
\end{figure*}

To quantify the physical and chemical properties of the L1517 cores,
we modeled their emission following the
procedure described in \citet{TAF04} for the analysis of
the L1498 and L1517B cores. In this way,
we assumed that the cores are 
spherically symmetric, and we concentrated our modeling 
effort on fitting the radial profiles of emission shown in 
Fig.~\ref{cores_radial}. To determine the density profile of each core,
we fitted the 1.2~mm continuum emission, as this is expected
to be the most faithful 
tracer of the total dust and gas column densities
(e.g., \citealt{BER07}).
Following the analysis of L1517B, 
we assumed a uniform dust temperature of $T_d=10$~K
and a 1.2mm emissivity of $\kappa=0.005$~cm$^2$~g$^{-1}$,
and we fitted the continuum radial profiles
with density laws of 
the form $n(r) = n_0/(1+(r/r_0)^\alpha)$, 
where $n_0$, $r_0$, and $\alpha$ are free parameters.
(For core B, we have used the \citealt{TAF04} result,
while core D was not fitted for lack of continuum data.)
The results of these fits are illustrated in Fig.~\ref{cores_radial},
and the best-fit parameters are 
summarized in Table~\ref{cores_fits}. As can be seen, core central 
densities range from $4.7 \times  10^4$~cm$^{-3}$ in core C
to $2.2 \times  10^5$~cm$^{-3}$ in core B,
which correspond, respectively, to
enhancements of 6 and 30 with respect to
the central density of the filaments traced in 
C$^{18}$O (section \ref{fils_density}). 
Integrating the density profiles
up to a representative radius of 0.05~pc ($\approx 75''$),
we estimate that the core masses are on the order of 1-2~M$_\odot$,
which is typical of the population of
Taurus starless cores \citep{BEN89}.

Once the density profile of each core had been modeled, we could determine
the abundance of the different species by solving the equation
of radiative transfer and convolving the result 
with the appropriate gaussian beam size to simulate an
observation. We did this using
a Monte Carlo non-LTE code based on that of 
\citet{BER79}, and we 
excluded core D from the analysis 
due to the lack of mm-continuum data.
For all the cores, we assumed 
a gas kinetic temperature of 9.5~K 
based on the
NH$_3$ analysis of L1517B in \citet{TAF04}, and in
agreement with
typical estimates for other low-mass starless cores \citep{BEN89}.
To reproduce
the typical N$_2$H$^+$(1--0) FWHM of 0.21~km~s$^{-1}$
(mean of values for all cores, see Table~\ref{cores_prop}), 
we added a
constant nonthermal component of 0.17~km~s$^{-1}$
(FWHM) to the velocity field.
Also, we extended each core density profile with a cloud
component based on the profile of the filament in which it is embedded.
The assignment of cores to filaments was done using
the core LSR velocity determined from N$_2$H$^+$ (Table~\ref{cores_prop}) 
and the velocity limits of each filament measured from
C$^{18}$O (Table~\ref{filaments}). According to this
criterion, cores A2 and C belong to filament 1, cores A1 and B
to filament 2, and core D to filament 3 
(as summarized in Table~\ref{filaments}). 
As discussed in section~\ref{fils}, there is some ambiguity
in the assignment of gas to filaments in the region where filaments
1 and 2 overlap, and this may affect the filament assignment for
cores A1 and A2. Our choice denotes our best understanding
of the emission 
after a careful inspection of the individual spectra from both
the FCRAO and IRAM 30m telescopes.

The above parametrization fixes
the physical structure of
each core and leaves the abundance law
of each molecular species
 as the only free
parameter left to fit the observed radial profiles of intensity.
Following previous work,
we explored two types of abundance laws.
As a first choice, we 
asumed that the molecular abundance is uniform across 
the core, and we used the Monte Carlo model
to predict the intensity radial profile.
Sometimes, this uniform abundance law predicted
an emission profile that is much steeper than observed,
and in these cases we modified the abundance law
to include   a central depletion hole 
of radius $r_d$ within which the abundance
is negligible.

\begin{table*}
\caption[]{Density and abundance profiles in the L1517 cores.$^{(1)}$
\label{cores_fits}}
\begin{tabular}{c|ccc|ccccc}
\hline
\hline
\noalign{\smallskip}
& \multicolumn{3}{|c|}{Density parameters} &
\multicolumn{5}{c} {Molecular abundances and depletion radii} \\
\noalign{\smallskip}
\hline
\noalign{\smallskip}
 Core  & $n_\circ$ & r$_{1/2}$ & $\alpha$ & X(N$_2$H$^+$) & X(SO) & r$_d$(SO) & X(C$^{18}$O) & r$_d$(C$^{18}$O) \\
       &   (cm$^{-3}$) & (arcsec) &    &           &           & (cm)  &                  & (cm)  \\
\noalign{\smallskip}
\hline
\noalign{\smallskip}
A1 & 7.0$\times$10$^4$ & 60 & 2.5 & 1.0$\times$10$^{-10}$ & 1.0$\times$10$^{-9}$ & 1.3$\times$10$^{17}$ & 1.5$\times$10$^{-7}$ & 1.5$\times$10$^{17}$  \\
A2 & 6.0$\times$10$^4$ & 60 & 3.5 & 0.4$\times$10$^{-10}$ & 1.0$\times$10$^{-9}$ & 0.0                   & 1.5$\times$10$^{-7}$ & 1.1$\times$10$^{17}$  \\
B & 2.2$\times$10$^5$ & 35 & 3.5  & 1.5$\times$10$^{-10}$ & 0.4$\times$10$^{-9}$ & 1.2$\times$10$^{17}$ & 1.5$\times$10$^{-7}$ & 1.9$\times$10$^{17}$  \\
C & 4.7$\times$10$^4$ & 60 & 2.5   & 0.7$\times$10$^{-10}$ & 2.0$\times$10$^{-9}$ & 0.0                   & 1.5$\times$10$^{-7}$ & 1.7$\times$10$^{17}$  \\
\hline
\end{tabular}
\begin{list}{}{}
\item[Notes:] (1) f All abundances are relative to H$_2$.
Core D was not modeled due to lack of mm-continuum data.
\end{list}
\end{table*}

As Fig.~\ref{cores_radial} shows, the N$_2$H$^+$ radial profile
of all the cores can be fitted using a uniform abundance model.
Each core, however, requires a different abundance value.
Cores A2 and C present the lowest 
N$_2$H$^+$ abundances, with best-fit values that are 
four and two times lower
than the value in core B, the most N$_2$H$^+$-rich core in the
sample (see Table~\ref{cores_fits} for absolute abundance values). 
As mentioned above, the N$_2$H$^+$ abundance in a core is expected to 
increase with time, so
we can interpret the observed differences in  N$_2$H$^+$ abundance
as differences in the state of the chemical evolution of
the cores. To better illustrate this evolution, we have 
re-ordered the cores in Fig.~\ref{cores_radial} 
so that the N$_2$H$^+$ abundance decreases  
towards the right. If our understanding of the 
N$_2$H$^+$ chemistry is correct, the sequence 
in the figure must represent
a sequence of cores of decreasing age,
in which core B is the oldest and cores A2 and C
are the youngest. 
(Core D has been placed in the sequence by considering
its N$_2$H$^+$(1--0) intensity towards the center, as no abundance
estimate was possible without a mm-continuum map.)

The analysis of the SO abundance provides further support to the
N$_2$H$^+$-derived age sequence.
As mentioned above, SO is expected to behave
opposite to N$_2$H$^+$, and therefore show the largest abundances in 
the chemically 
youngest cores. Indeed, Fig.~\ref{cores_radial} and Table~\ref{cores_fits}
show that the young cores A2 and C have large and
uniform SO abundances, while the more evolved B and A1 cores
present SO depletion holes towards their center. 

Finally, all cores require a depletion hole
in their C$^{18}$O abundance law to 
fit the observed emission profile.
The youngest A2 core requires the smallest hole, while the oldest
B core requires the largest one (Table~\ref{cores_fits}). 
Still, the correlation between
evolutionary stage and C$^{18}$O depletion hole is weak,
probably because 
determining the correct C$^{18}$O abundance law
depends on modeling
the poorly constrained envelope that surrounds each core.
In any case, the results of the C$^{18}$O analysis
are consistent with the interpretation that the L1517 cores 
are at different stages of evolution.

Before finishing our analysis, it
is worth mentioning that in the evolutionary
sequence of Fig.~\ref{cores_radial},
core D, for which no modeling could
be carried out, appears as a relatively evolved core
based both on its strong N$_2$H$^+$ emission
and on the flattened radial profiles of
SO and C$^{18}$O.
This conclusion, however, disagrees
with the recent analysis by \citet{HIR09}, who suggests
that L1517D is an unusually young core from
a chemical point of view. 
Resolving this contradiction requires making
mm-continuum observations of this object to
determine its true density structure and to allow 
a Monte Carlo modeling of its abundance profiles
like that of the other cores.

Although the cores in L1517 differ significantly in their
chemical composition and therefore evolutionary state, 
they present very similar kinematic properties.
Table~\ref{cores_prop} shows that the N$_2$H$^+$ linwewidths 
are all close to 0.2~km~s$^{-1}$, which is typical of the 
low-mass cores in Taurus and indicative
of subsonic nonthermal motions \citep{MYE83_2,CAS02}. 
In addition, the 
internal velocity gradients of the cores are
on the order of 1~km~s$^{-1}$~pc$^{-1}$, which is also typical
of low-mass cores \citep{GOO93,CAS02}, and again suggests that 
the L1517 cores are representative of the population
of low-mass cores in Taurus as a whole.
This combination of different chemical composition and 
similar internal velocity structure suggests that the
kinematic properties of the cores change little
during their contraction from the more
diffuse gas in the cloud. To further elucidate this issue, 
in the next sections we analyze the kinematics of the more
extended gas in the cloud, and we compare it with
that of the cores.

\section{Gaussian decomposition of the spectra}\label{gaus}

\begin{figure}
\centering
\resizebox{\hsize}{!}{\includegraphics{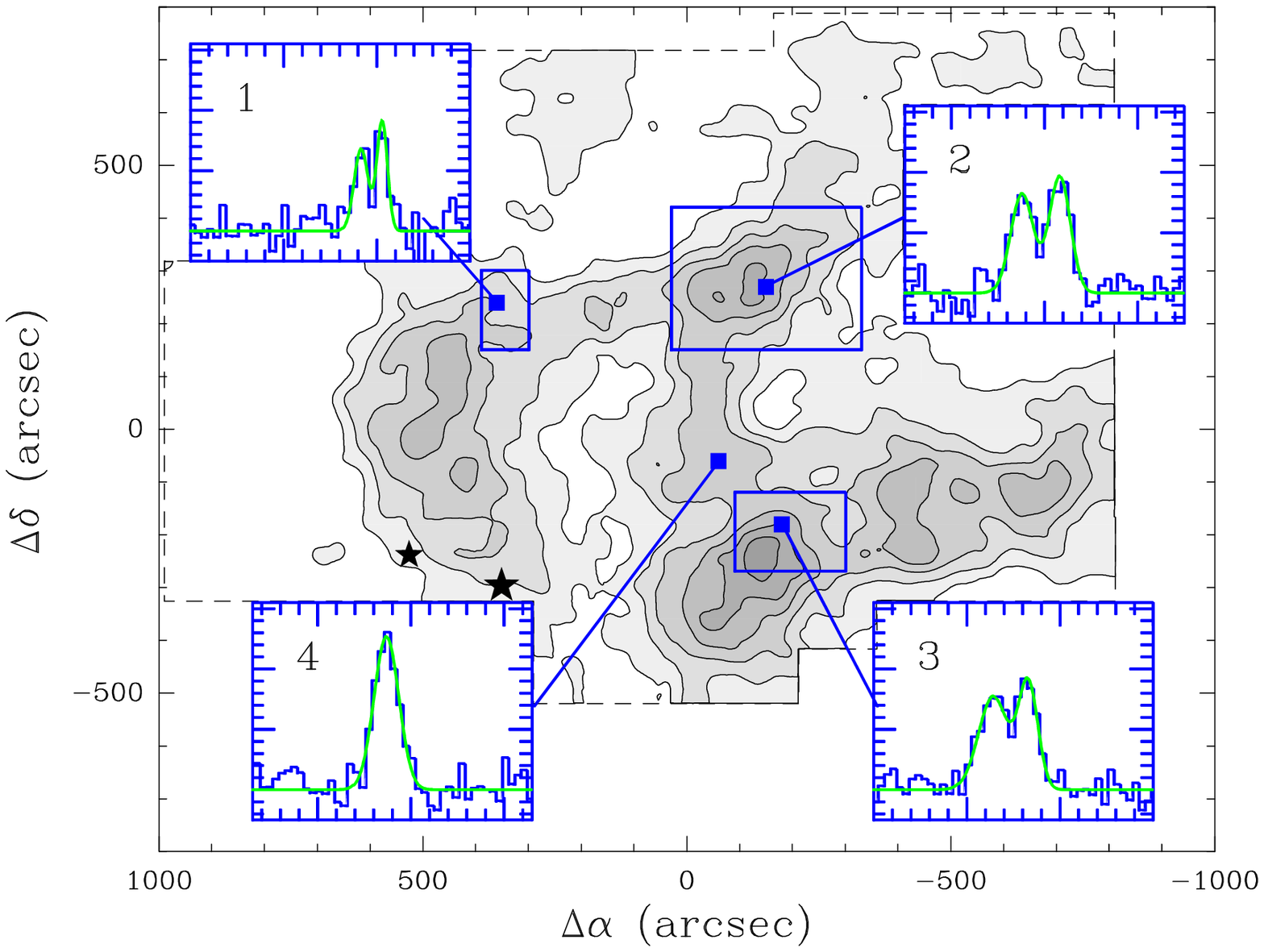}}
\caption{Graphical summary of the Gaussian fits to the 
C$^{18}$O(1--0) emission in L1517. 
The  gray scale shows the C$^{18}$O(1--0)
integrated intensity map of  Fig.~\ref{DSS_C18O}, and the superposed
blue boxes enclose the regions where two Gaussians were fitted to the
spectra. Insets with numbers 1, 2, and 3 illustrate double-Gaussian
fits, while inset number 4 shows an example of the dominant
single Gaussian fitted to most of the cloud spectra.
\label{double}}
\end{figure}

In section \ref{fils} we saw 
that each filament in the
L1517 cloud is characterized by a well-defined range of LSR velocities 
(Table~\ref{filaments}). 
The filaments, however, are not monolithic entities and
seem to have an internal velocity structure characterized by 
low-level velocity gradients. As these gradients
have potential information on the 
gas motions responsible for forming
the embedded dense cores, we analyzed them in more detail
by inspecting the individual C$^{18}$O(1--0) spectra. 
From this inspection, we find that
92\% of the spectra with peak intensity S/N$ \ge 3$
present a single velocity component, while the
rest of the spectra present evidence of
two partly-overlapping velocity components. This preponderance of
profiles with a well-defined number of components 
suggests that the velocity field 
at each cloud position can be characterized
by fitting Gaussians to the spectra.

To fit the more than 2,000
C$^{18}$O(1--0) spectra 
observed towards L1517
we have used a semi-automatic procedure in the 
CLASS software and chosen to fit
one or two components depending on the complexity of the line profile.
The result from this fitting is illustrated
in Fig.\ref{double}, which also shows
examples of the profiles.
As can be seen, most of the cloud spectra have been fitted with single Gaussians
(as shown in box number 4), and the fits with two Gaussians are
restricted to  three well-defined
regions identified with solid boxes.  
The largest of these three regions (box number 2) is centered
near  $(\Delta \alpha, \Delta \delta)=$ (-200$''$, +250$''$) and seems to
correspond to the superposition of filaments 3 and 4.
This region appears in optical images as a relative enhancement
in the obscuration, and \citet{LEE99a} even classified it as
an additional core (L1517B-2), but 
our observations suggest that
it is more likely a superposition effect.
Another region with double spectra (box number 1) also seems to result from the
superposition of two different components, this time filament 3 and an unlabeled
and more diffuse component parallel to filament 4 that can be seen in the 
reddest map of 
Fig.~\ref{C18O_channelmap}. 
Finally, the third region with double spectra  occurs
near $(\Delta \alpha, \Delta \delta)=$ (-200$''$, -200$''$) and has a 
less clear origin. Its location would suggest  an origin in 
the superposition of filaments 1 and 2, but the velocity of the 
blue component does not match the velocity of any of the filaments.
Observations with higher S/N towards this region are needed to 
determine the origin of this anomalous component.

The Gaussian fitting procedure has also been
applied to the SO and N$_2$H$^+$ spectra. For SO, which is almost as
extended spatially as C$^{18}$O (see Fig.~\ref{densegas}), we again find that
we need to use two Gaussian components
inside the regions discussed before,
while single-peaked spectra are
the norm in the rest of the cloud.
For the N$_2$H$^+$(1--0) spectra,
we fitted all hyperfine components simultaneously 
using the HFS method in CLASS and derived 
both the line centroid and the optical-depth corrected 
linewidth in this way. As shown in Fig.~\ref{densegas}, the  N$_2$H$^+$ emission
is much more compact than the C$^{18}$O and SO emission,
and our inspection of the individual spectra 
found that all positions could be fitted with
a single velocity component.

The linewidths and velocity centroids derived with the Gaussian
fits just described  constitute the two input parameters for
our analysis of the gas kinematics in L1517. In the
following two sections, we analyze the
behavior of each quantity separately.

\section{Linewidth analysis}

\subsection{Linewidth statistics: prevalence of subsonic 
motions}\label{linewidths1}

The linewidth of a spectrum combines contributions
from both thermal and nonthermal gas motions. 
In this section, we are interested 
in the nonthermal motions of the gas, which are potentially
associated to gas
turbulence or core formation motions.
To isolate these motions,
we have subtracted the thermal component
from the measured linewidth
following the standard practice of assuming that
the two contributions are independent of
each other so they add in
quadrature (e.g., \citealt{MYE83_2}). In this way, we
estimate the nonthermal velocity dispersion for each species
as
\begin{equation}
\label{sigma_nt}
\sigma_{NT} = \sqrt{\frac{\Delta V^2}{8 \ln2} -
\frac{k T}{m}},
\end{equation}
where $\Delta V$ is the measured FWHM linewidth, 
$k$ the Boltzmann's constant, $T$ the gas kinetic temperature,
and $m$ the mass of the molecule under consideration.
This velocity dispersion $\sigma_{NT}$ can be directly
compared to the (isothermal)
sound speed of the gas, $c_s$, which 
has a value of 
0.19~km~s$^{-1}$ for ISM gas at 10~K.

\begin{figure}
\centering
\resizebox{\hsize}{!}{\includegraphics{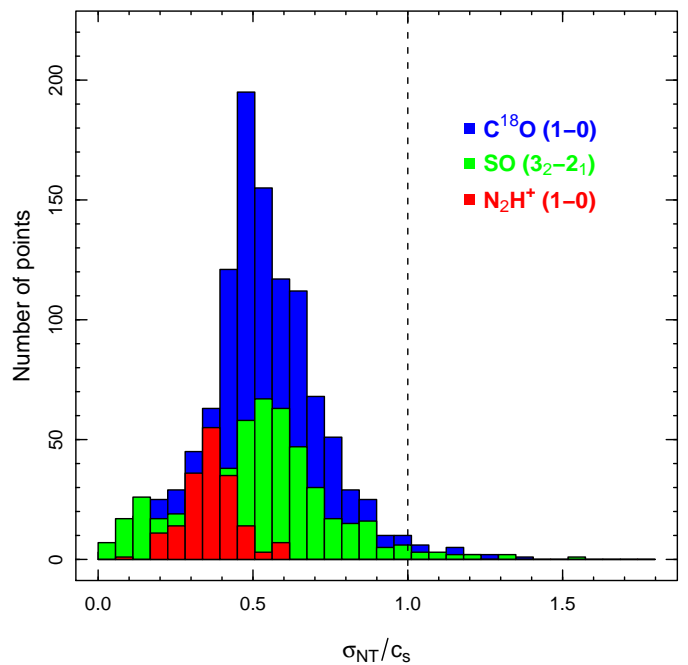}}
\caption{Histogram of the nonthermal velocity dispersion for 
C$^{18}$O (blue), SO (green), and N$_2$H$^+$
(red) illustrating the predominance of subsonic values.
\label{DVNth}}
\end{figure}

Figure~\ref{DVNth} shows the histograms of the velocity 
dispersion for the three molecular species
in our survey  (using all spectra with peak intensity 
S/N$ \ge  3$).
As can be seen, the overwhelming majority of 
the spectra present a subsonic velocity dispersion
and lie to the left of the 
$\sigma_{NT}/c_s = 1$ dashed line.
The few positions that exceed the sonic limit 
form a low-level tail in the distribution that
contains 2\% of the spectra in C$^{18}$O
and 3\% of the spectra in SO. No N$_2$H$^+$
spectrum  has a supersonic linewidth.

In addition to a very small fraction of supersonic 
points, the
main feature of the histograms in Fig.~\ref{DVNth} is
the presence of
a well-defined central peak. This peak indicates 
that the gas in the cloud is not only subsonic, but
 it also has
a favored number of nonthermal motions that 
approximately equal half the sound speed. 
More specifically, the histograms in the figure
have the following mean and rms:
$\sigma_{NT}/c_s = 0.54 \pm 0.19$ for C$^{18}$O,
$\sigma_{NT}/c_s = 0.51 \pm 0.23$ for SO, and
$\sigma_{NT}/c_s = 0.36 \pm 0.09$ for N$_2$H$^+$.

The values for the C$^{18}$O and SO nonthermal
linewidth seem consistent with each other.
The mean $\sigma_{NT}/c_s$ of N$_2$H$^+$, 
on the other hand, is significantly smaller than that
of C$^{18}$O and SO. Optical depth effects could
potentially play a role in this difference, since the hyperfine fit to
the N$_2$H$^+$(1--0) spectrum automatically
corrects for optical depth broadening, 
while no correction is 
applied to the C$^{18}$O and SO data when fitting
single Gaussians. To test whether optical depth
broadening
can explain the higher C$^{18}$O mean value, 
we have applied 
a simple correction to the C$^{18}$O data by assuming 
an excitation temperature of 10~K
and estimating the optical depth from the
peak intensity of the observed line profile. 
This correction is relatively small, as
the optical depth never reaches unity, so even
after its application, 
the distribution of 
$\sigma_{NT}$ still has a significantly higher mean value 
than for N$_2$H$^+$, and
both a Kolmogorov-Smirnov and a 
Wilcoxon-Mann-Whitney test confirm that 
the two distributions must be different.

The lower  $\sigma_{NT}$ value
of the  N$_2$H$^+$  spectra
could indicate that the dense core
gas has an intrinsically lower velocity dispersion
than the less-dense ambient material 
and that core formation has been accompanied by certain 
amount of dissipation of nonthermal
motions. 
If this is the case,
we can quantify the loss of nonthermal motions 
from the difference 
between the FWHM linewidths of C$^{18}$O and N$_2$H$^+$,
which is  
$0.07\pm 0.04$ km~s$^{-1}$.

Even if small, the above estimate 
of the loss of non thermal motions during core formation
is likely a significant overestimate.
The nonthermal
linewidth is not a local parameter, but the accumulated
effect of gas motions along the line of sight. For
this reason, 
some consideration should be given to line-of-sight effects
when comparing tracers as different as C$^{18}$O and N$_2$H$^+$.
As shown in section \ref{core_model}, the
N$_2$H$^+$ emission is significantly more concentrated 
than the emission from C$^{18}$O and SO 
because of its special chemistry 
and excitation requirements, so it samples a 
shorter line of sight path than C$^{18}$O and SO.
In addition, we see in the next section that part of 
the kinematics of the C$^{18}$O/SO-emitting gas 
arises from large-scale motions in the filaments.
As these motions likely accumulate along the line of sight,
they are expected to contribute more to the C$^{18}$O and SO
linewidth than to N$_2$H$^+$.

To estimate the importance of this effect, we compare the
sizes of the typical emitting regions for N$_2$H$^+$
and C$^{18}$O. 
Tables~\ref{Fils_fits} and \ref{cores_fits} show that
the typical diameter of an L1517 core traced in N$_2$H$^+$
is about $2 \times 55''$, while
the same parameter for a filament traced in C$^{18}$O
is about $2\times 95''$. 
In addition, 
Table~\ref{fil_kinematics} shows that the typical
velocity gradient in the filament gas 
is about 1.2~km~s$^{-1}$~pc$^{-1}$. Thus, if we
assume a gradient of this size and multiply it
by the difference in line of sight length 
between the two molecules, we predict a 
velocity difference between the tracers 
due to large-scale motions of
about 0.06~km~s$^{-1}$. 
This difference is so close to the observed
linewidth difference
that it seems very likely that a significant 
part of the nonthermal
linewidth difference between N$_2$H$^+$ and C$^{18}$O 
(and SO) arises from their
different sampling of the cloud
large-scale motions.
Whether this effect can explain 
the full linewidth difference 
cannot be said, but it does show that the
difference in local nonthermal motions (and
thus dissipation during core formation) must
be necessarily less than estimated before.

A simple consequence of the subsonic nature of the
nonthermal gas motions in L1517 is the small role that
they can play in supporting the cloud against gravity.
The ratio between the nonthermal and thermal 
contributions to the gas pressure is given by
$P_{NT}/P_T = (\sigma_{NT}/c_s)^2,$ so even if we use
the  C$^{18}$O
linewidth uncorrected for optical depth, we find that 
the nonthermal motions contribute to the gas
pressure with only $\sim 1/3$ of the thermal value.
This contribution is likely to represent an upper limit,
because the above pressure ratio assumes again 
that all nonthermal motions
arise from a local ``microscopic'' turbulent motions that increases cloud
support \citep{CHA51} and ignores the contribution from
large-scale flows.

\begin{figure}
\centering
\resizebox{\hsize}{!}{\includegraphics{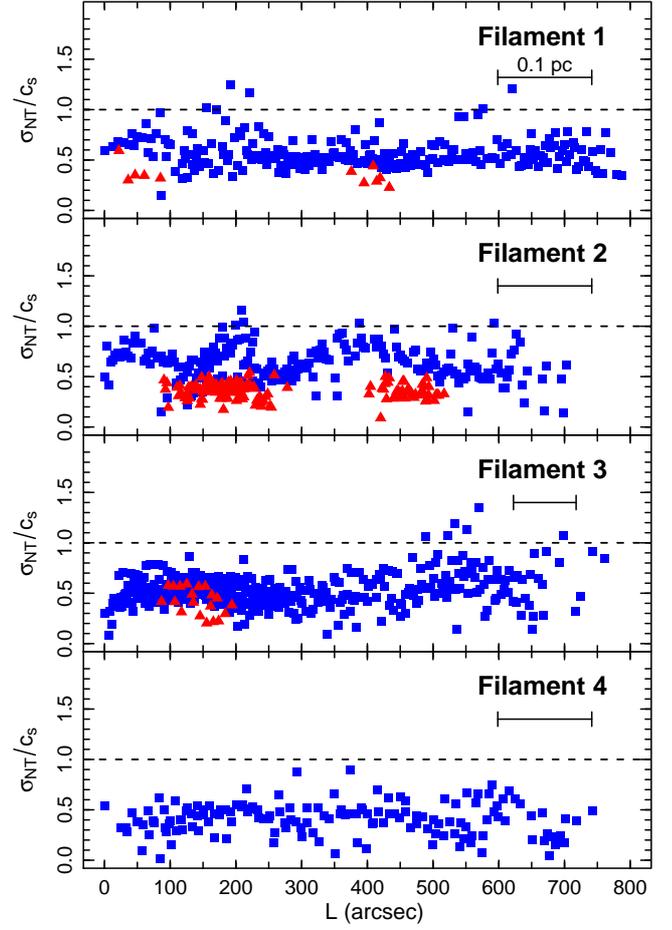}}
\caption{Distribution of nonthermal velocity dispersion as a function
of position along each of the L1517 filaments.
Blue squares represent C$^{18}$O(1--0) and red triangles 
and N$_2$H$^+$(1--0) values.
The spatial scale for filament 3 has been shrunk by a factor 
of 1.5 in order to fit it into the reduced box size used for
the other filaments.
The mean formal error in the $\sigma_{NT}/c_s$ determination is 0.1 for both N$_2$H$^+$ and C$^{18}$O, which is approximately the marker size.
\label{DVNth_fil}}
\end{figure}

\subsection{Spatial distribution and consequences
for turbulent models of core formation}\label{linewidths2}

To further characterize the nonthermal 
motions, we now study their spatial distribution. 
Figure~\ref{DVNth_fil} presents a plot of
$\sigma_{NT}/c_s$ for C$^{18}$O 
 and N$_2$H$^+$ 
as a function of position along
each of the filaments identified in section~\ref{fils}.
In agreement with the histogram analysis, 
all four filaments are dominated by subsonic gas 
over their entire length, and 
only a few positions exceed the sonic limit
in C$^{18}$O.
These few supersonic positions 
(some of them coincident with filament overlaps)
appear almost randomly
scattered over the filaments length,
and show no correlation with either the position
of the cores 
or the beginning or end of the filaments.

In addition to a lack of supersonic points, the plots
of nonthermal linewidth in Fig.~\ref{DVNth_fil}
present very little spatial structure.
Filaments 1, 3, and 4 have almost flat distributions of 
C$^{18}$O
$\sigma_{NT}/c$ and well-defined mean values close to 
0.5 (Table~\ref{fil_kinematics}).
Filament number 2 also has a mean
$\sigma_{NT}/c$ close
to 0.5, but presents changes in the linewidth
along its length. These changes are of unclear origin,
and seem to result from the presence of two localized 
regions of enhanced (but mostly subsonic) linewidth. One of these
regions corresponds to the vicinity of
core B, and the other occurs in the vicinity of core A1.
Concerning the N$_2$H$^+$ linewidths, Fig.~\ref{DVNth_fil}
again shows their tendency to be 
smaller than those of C$^{18}$O. This trend is present in
all the cores, although it is somewhat reduced in core D of filament 3.
Apart from this already discussed effect, Fig.~\ref{DVNth_fil}
shows that the N$_2$H$^+$ linewidths remain approximately constant 
inside each of the cores. 

\begin{table}
\caption[]{Kinematic properties of the filaments$^{(1)}$
\label{fil_kinematics}}
\begin{tabular}{cccc}
\hline
\hline
\noalign{\smallskip}
 Filament  & $\langle \sigma_{NT}\rangle/c_s$ & $\langle
V_{LSR}\rangle$  & $\langle|\nabla V_{LSR}|\rangle$  \\
      &   & (km s$^{-1}$) & (km s$^{-1}$pc$^{-1}$) \\
\noalign{\smallskip}
\hline
\noalign{\smallskip}
1  & 0.57$\pm$0.15 & 5.52$\pm$0.07  & 1.0$\pm$0.5 \\
2  & 0.63$\pm$0.17 & 5.79$\pm$0.07  & 1.4$\pm$0.7 \\
3  & 0.53$\pm$0.16 & 5.89$\pm$0.05  & 1.3$\pm$1.0 \\
4  & 0.41$\pm$0.16 & 6.14$\pm$0.04  & 0.9$\pm$0.5 \\
\hline
\end{tabular}
\begin{list}{}{}
\item[Notes:] (1) From C$^{18}$O(1--0) emission.
\end{list}
\end{table}

\begin{figure}
\centering
\resizebox{\hsize}{!}{\includegraphics{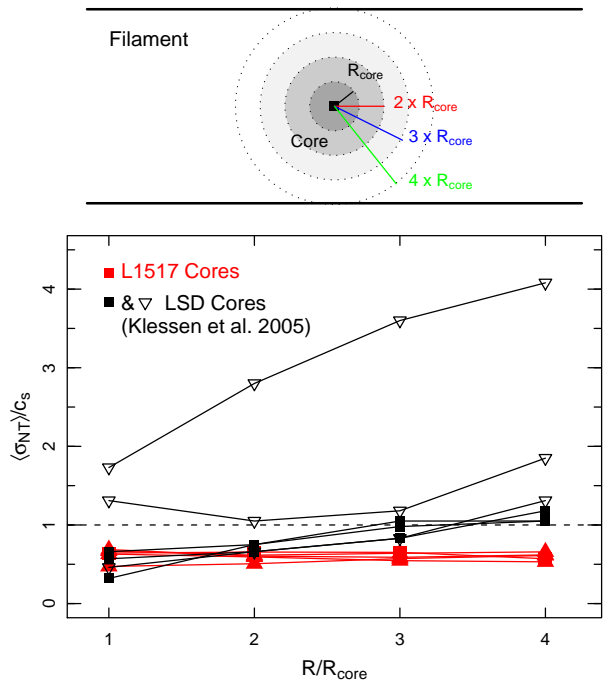}}
\caption{Comparison between the C$^{18}$O nonthermal velocity dispersion
around each L1517 core and the
predictions from a model of turbulent core formation. {\bf Top: }
schematic view illustrating how the velocity dispersion has been 
averaged inside
four concentric shells around each core. {\bf Bottom:} comparison
between the observed C$^{18}$O data (red solid triangles) and
the predictions from the LSD model by \citet{KLE05}. The 
black solid squares represent the three different projections of
a fully subsonic core, while the open inverted triangles 
represent the three projections of a core that has 
transonic motions in one of them.
\label{sigmaNth_cores}}
\end{figure}

The flat distribution of the velocity dispersion in the filaments, and
in particular in the vicinity
of the dense cores, poses a significant constraint to models
of core formation by supersonic turbulence.
In these models, the cores 
are shock-compressed structures formed at the stagnation
point between two convergent flows, and as a result, 
they are expected to be  
surrounded (even confined) by turbulent layers of cloud gas (e.g.,
\citealt{PAD01,BAL03,KLE05}). In L1517, the observed C$^{18}$O emission 
surrounds the location of the dense cores and arises from gas that is
about one order of magnitude less dense than the core 
centers (section~\ref{core_model}), 
so we can naturally expect that the C$^{18}$O emission
arises from the gas layers that physically
surround the dense cores. (C$^{18}$O does
not sample gas inside the cores due to freeze out, see 
section~\ref{cores}.) If the cores have formed by the
convergence of gas flows, 
part of the C$^{18}$O emission must 
originate in gas in the flows and must retain a 
signature from the core-forming 
convergent motions. To test whether such a signature
is present in our data, we calculated for each core
the average C$^{18}$O velocity dispersion inside rings 
around the core center having radii equal to 1, 2, 3, and 4
core radii (core radius is $r_{1/2}$ in Table~\ref{cores_fits}).
This procedure is illustrated in the upper panel
of Fig.~\ref{sigmaNth_cores}, and has been stopped
at four core radii to ensure that we sample equally
gas that surrounds the cores in
all directions, and not only in the direction of the large-scale
filaments (which dominates the
emission at larger radii). The results from this average 
for each of the five cores in L1517 are 
indicated in the bottom panel of 
Fig.~\ref{sigmaNth_cores}. Not surprisingly given
the general trend seen in the filaments, all cores
present a distribution of velocity dispersion with radius that 
has no systematic change with distance
from the core center.

To compare the distribution of velocity dispersions  
around the cores of L1517
with the predictions from turbulent models, we chose the numerical simulations of 
gravo-turbulent core formation by \citet{KLE05}.
These authors attempt to reproduce conditions
similar to those
of a low-mass star-forming region like L1517
and present a very complete view of the
internal kinematics of the simulated cores. 
From the two families of simulations that these authors
present, we selected the large-scale driving (LSD) case
because it produces the most quiescent cores, and is
therefore more likely to fit our observations of L1517.
\citet{KLE05} present in their Fig.~1
radial profiles of the velocity dispersion for three different views of
two selected cores (likely chosen for their similarity to
observations), and we used these profiles to derive
nonthermal velocity components 
by subtracting in quadrature the contribution 
from thermal motions
(as done with the C$^{18}$O data). These data have been 
averaged as a function of core radius by converting the 
normalized column densities given by the authors
into a distance from the core center by assuming that 
the the density of the cores follows the profile of a
Bonnor-Ebert sphere (as claimed by the authors). 

The predicted nonthermal velocity dispersions 
from the \citet{KLE05} model are represented
in Fig.~\ref{sigmaNth_cores}.
As expected, these predicted dispersions 
show a significant increase with radius due to
the presence of converging flows around the cores.
The predicted increase is
approximately a factor of 2 between 
1 and 4 core radii, and as a result, the outermost points 
in the graph are expected to exceed the sound speed limit
in all core models.
Even in the most quiescent, 
subsonic core model, 
which is
arguably not representative of the sample because subsonic
cores are less than 25\% of the total in the simulation, 
the match between model and data is only acceptable for the 
innermost two radii. For larger radii, model and data 
diverge with increasing distance from the core center,
and end up differing by a factor of 2 in the outermost layers
sampled by our observations. Even larger disagreement
occurs for the case of transonic cores,
which constitute $\sim 50$\% of the total
cores in the simulation and are therefore more
representative of the model results. As can be seen,
two out of the three sets of points exceed the sound
speed at all radii, and therefore fail to fit 
the observations at all radii in all the cores.

Although limited, 
the above comparison illustrates the basic 
disagreement between our observations of the L1517 cloud
and the predictions from models of core formation
by convergence of supersonic flows:
the C$^{18}$O spectra in L1517 sample the core outer
layers far enough to trace any core-forming motions, 
so if these motions were supersonic,
they should have left a clear signature in the C$^{18}$O
spectra.
The absence of such a signature 
rules out the presence supersonic motions around the cores
and sets a limit to 
any core-forming gas flow that is 
well inside the subsonic regime.

\subsection{Velocity coherent filaments}

In their study of 
the linewidth of dense-gas
tracers in cores,
\citet{BAR98} and \citet{GOO98} found
that the nonthermal component 
remains almost constant over 
the core interior and that this behavior represents a departure
from the well-known 
linewidth-size relation commonly associated to 
turbulent motions \citep{LAR81}.
These authors refer to this property
of the core gas as velocity ``coherence,'' and suggest that 
the scale of coherence, which they found to 
approximately coincide with the
core size, may indicate that core formation is 
related to the process of 
turbulence dissipation.

Our analysis of the C$^{18}$O linewidth
illustrated in 
Figs.~\ref{DVNth_fil} and \ref{sigmaNth_cores}
shows 
that the region with constant nonthermal linewidth
can be followed in significantly larger scales 
than the cores
and that it extends over distances as long
as the filaments themselves ($\sim 0.5$~pc).
This coherence of the velocity field in the filaments
is not limited to the nonthermal linewidth, and
it can be seen in the behavior of the velocity centroid,
to be studied in detail in the next section. 
In the L1517 cloud, therefore, the filaments
are velocity coherent in the sense of 
\citet{BAR98} and \citet{GOO98}, and 
this extends the scale size of coherence in the gas
by a factor of approximately 5.

Finding velocity coherence on scales as
large as the filaments 
separates the scale of coherence
from that of the cores, and suggests that 
core formation is not likely the direct result of
turbulence dissipation. 
As seen in section~\ref{linewidths1}, some
turbulence dissipation may have occurred during
core formation, but
by being significantly subsonic, 
it is unlikely to have affected the
gas pressure balance and
triggered core formation.
Velocity coherence in cores, therefore, 
does not seem a defining property of the 
condensations, but a condition
inherited from the larger spatial scale
of the filaments.

Our finding of velocity coherence on scales larger than 
a core would seem to contradict
the recent results from 
\citet{PIN10}, who claim to have
detected of the transition between the coherent
and turbulent regimes in a core with observations 
of the B5 region in Perseus.
We note, however, that although these
authors associate the velocity coherent
region with a dense core, 
one can see in their Fig.~3 that
the velocity coherent region  in B5
is elongated and has a length of
about 0.5~pc, similar to our L1517 filaments.
The B5 region, in addition, contains multiple
dense cores (\citealt{ENO06}, also our own 
unpublished data), and this suggests again
that it is more similar to the L1517 filaments 
than to an isolated dense core.
Clearly more observations of different core
environments are needed to understand
the size and relation between the different 
velocity coherent regions. If the L1517 cloud
is a representative region of dense core formation,
we can predict that filament-wide velocity coherence
will be a common phenomenon.

\section{Line centroid analysis}
\label{linecent}

\begin{figure}
\centering
\resizebox{\hsize}{!}{\includegraphics{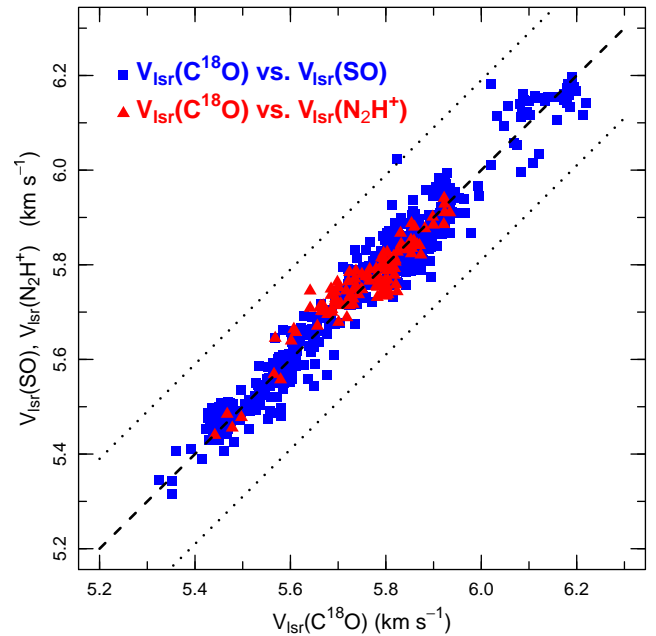}}
\caption{Comparison between the $V_{LSR}$ of C$^{18}$O and SO (blue
squares), and C$^{18}$O and N$_2$H$^+$ (red triangles) in L1517
illustrating the good agreement between all tracers. The central dashed 
line indicates the locus of equal velocities, and the surrounding 
dotted lines have been displaced by the sound speed (0.19~km~s$^{-1}$
for ISM gas at 10~K).
The mean formal error in the $V_{LSR}$ determination is 0.02~km~s$^{-1}$ for SO and 0.01~km~s$^{-1}$ for both C$^{18}$O and N$_2$H$^+$, 
which is comparable or smaller than the marker size.
\label{vel_tracers}}
\end{figure}

The second output from our Gaussian fit to the spectra is the 
velocity centroid.
Figure~\ref{vel_tracers} compares the 
SO and N$_2$H$^+$ centroids with those of C$^{18}$O towards all positions
where the fits were considered to be significant (S/N $\ge 3$).
As can be seen, the velocities of the different tracers agree
with each other 
independently of any variations in the bulk velocity of the 
cloud. The average difference in velocity between C$^{18}$O and 
either SO or N$_2$H$^+$  is
0.03$\pm$0.03 km s$^{-1}$, which means that 
the velocities of the different tracers differ on
average by less than one fifth of the sound speed. 
This good match between tracers 
rules out any significant motions between 
the different density regimes of the gas and, in particular,
it rules out any systematic drift
between the dense cores (traced by N$_2$H$^+$) and the
surrounding gas (traced by C$^{18}$O).
Such a quiescent state of the gas is not peculiar to L1517, and \citet{WAL04}
and \citet{KIR_H07} have found a similar lack of velocity
shifts between N$_2$H$^+$ and C$^{18}$O in a number of low-mass
star-forming regions.

\begin{figure}
\centering
\resizebox{\hsize}{!}{\includegraphics{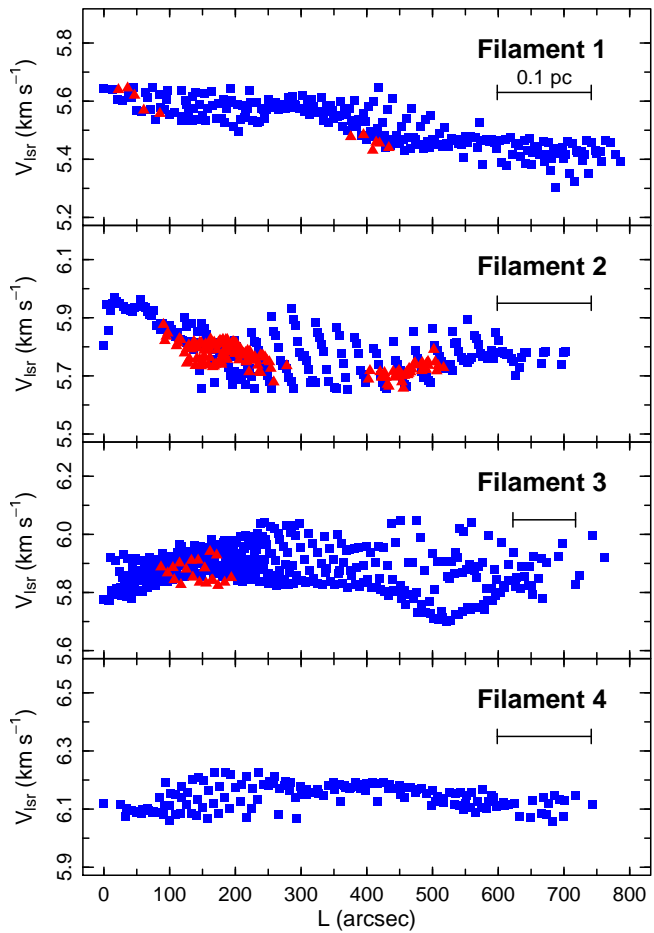}}
\caption{Velocity centroid as a function of position along each of
the L1517 filaments.
Blue squares represent C$^{18}$O(1--0)
and red triangles N$_2$H$^+$(1--0) values.
The spatial scale for filament 3 has been shrunk by a factor
of 1.5 in order to fit it into the reduced box size used for
the other filaments. 
The mean formal error in the velocity centroid determination is 0.02~km~s$^{-1}$ for C$^{18}$O and 0.01~km~s$^{-1}$ for N$_2$H$^+$, 
which is always smaller than the marker size.
\label{FilVlsr}}
\end{figure}

\subsection{Continuity of the velocity field and large-scale 
oscillations}\label{contin}

The lack of velocity shifts between tracers does not mean that the
gas in L1517 is static. Figure~\ref{vel_tracers} shows how the
LSR velocity of the cloud material spans almost 1~km~s$^{-1}$, and this
velocity spread
arises from a combination of the different velocities of each
of the filaments and the presence of 
internal velocity gradients inside the filaments. 
To more fully study these gradients, we
present plots of the velocity centroids of 
C$^{18}$O and N$_2$H$^+$ as a function of position along
each filament in Fig.~\ref{FilVlsr}. 

As can be seen in Fig.~\ref{FilVlsr}, the velocity field
of each filament presents a remarkably low level of 
spatial change. The
dispersion of the C$^{18}$O  velocity centroids
does not exceed 0.07~km~s$^{-1}$ in any filament,
even when considering a
few regions of enhanced dispersion that seem to
coincide with locations of multiple-peaked spectra and to
overlap between different filaments. In addition,
the end-to-end velocity change in each filament is about
0.2~km~s$^{-1}$ or less. This quiescent state of the
gas on scales as large as 0.5~pc  
once more justifies the interpretation of the filaments
as velocity-coherent structures. 

Although the changes in velocity along the filaments are 
small, the low dispersion of the centroid data allows
 the existence of large-scale velocity patterns 
to be discerned 
in some of the filaments. Figure~\ref{FilVlsr} shows how
the C$^{18}$O centroids oscillate quasi-periodically, 
especially in filaments 1 and 2,  
over the length of the filaments with typical 
wavelengths of  approximately 0.1-0.2~pc.
As also shown in the figure, the more spatially
localized N$_2$H$^+$ centroids
follow the C$^{18}$O oscillatory pattern at those positions where
both tracers can be observed (i.e., towards the dense cores),
indicating that both species 
trace the same general velocity pattern.
This good match between the C$^{18}$O  and N$_2$H$^+$
data is remarkable because the two species
do not coexist spatially due to their anticorrelated chemistries 
(section \ref{cores}), and they therefore
trace significantly different regimes of gas density.
Thus, the 
continuity of the C$^{18}$O  and N$_2$H$^+$ velocity 
gradients must reflect 
a continuity between the 
velocity field of the filament gas 
and the internal velocity gradients 
of the dense cores embedded in it. This implies 
that the internal velocity gradients in the cores
are not intrinsic core properties
(like isolated rotation),
but result from the large-scale motions of 
surrounding filament.
Core velocity gradients should therefore be
interpreted in terms of the velocity gradients of the
lower density gas.

The observed continuity between the velocity field of the filaments 
and the cores implies that core formation in L1517 has not decoupled
 the dense kinematically gas from its surrounding environment.
This behavior again seems to contradict the expectation
from the turbulent models of core formation, because in these models,
the cores are formed by the direct shock of streams of lower density gas, so a discontinuity in the velocity field is expected. 
The observations of L1517 are better understood if the filaments
are coherent in velocity as a whole, and the cores have
formed from the contraction of this velocity-coherent gas. 
The transition from cloud to core conditions 
must have therefore 
involved little or no dissipation of kinetic energy, and in 
particular the absence of supersonic shocks. The data therefore 
indicates that core formation
in L1517 has been an almost quasi-static process.

\subsection{A simple model of the velocity oscillations}\label{simple_model}

The velocity oscillations of filaments 1 and 2,
dominate the large-scale kinematics of the gas
in these objects, and 
suggest some type of ordered pattern
in their three-dimensional velocity field.
Our limited observations cannot fully constrain the
3D properties of this velocity field,
but the large-scale nature of the motions and the
lack of systematic velocity gradients
perpendicular to the filament axis
suggest that the underlying velocity field must have
an important component 
directed {\em along} the axis of the filaments.
If this is correct, it is intriguing to investigate 
whether the velocity oscillations of filaments 1 and 2 
are related to the presence of embedded cores, and
in particular, whether they could represent part
of the motions responsible for core formation.

Core formation in a filamentary gas cloud 
is often investigated
using the idealized geometry of an
infinitely-long, axially-symmetric gas cylinder. The stability
of such a configuration has been studied in detail
using both semi-analytical techniques 
\citep{STO63,OST64,LAR85,NAG87,INU92,NAK93,GEH96,FIE00a}
and numerical simulations \citep{BAS83,BAS91,NEL93,NAK95,FIE00b}. 
From this body of work we know
that, depending on the initial 
equilibrium state of the gas, the amount of support provided by pressure
and magnetic fields, and the geometry of both the magnetic field and
the perturbation
applied to the system, the gas cylinder can follow
a number of evolutionary paths, including expansion, collapse to a spindle,
or fragmentation into multiple clumps. Among these outcomes, 
cloud fragmentation is the most interesting here because of
its possible relation to core formation and 
the velocity oscillations in
filaments 1 and 2.

In the simplest case of an isothermal
cylinder in equilibrium, 
fragmentation occurs through a Jeans-type of 
instability in which a density perturbation 
with a large-enough wavelength accumulates
enough mass in each over-dense region
to make it gravitationally unstable 
(e.g., \citealt{STO63,LAR85}).
The most unstable perturbation 
is an axisymmetric mode that breaks
the initially continuous 
filament into a chain of dense condensations 
equally spaced along the cylinder axis \citep{NAK93}.
Numerical simulations show that this process
involves redistribution of the gas in the filament
via motions that
have a  dominant velocity component 
parallel to the filament axis, at least 
during the first stages of evolution
(e.g., \citealt{NAK93,FIE00b}). The
similarity of these motions with those
inferred for filaments 1 and 2 in L1517
is the main motivation for attempting 
a simple kinematic model.

\begin{figure*}
\centering
\resizebox{\hsize}{!}{\includegraphics{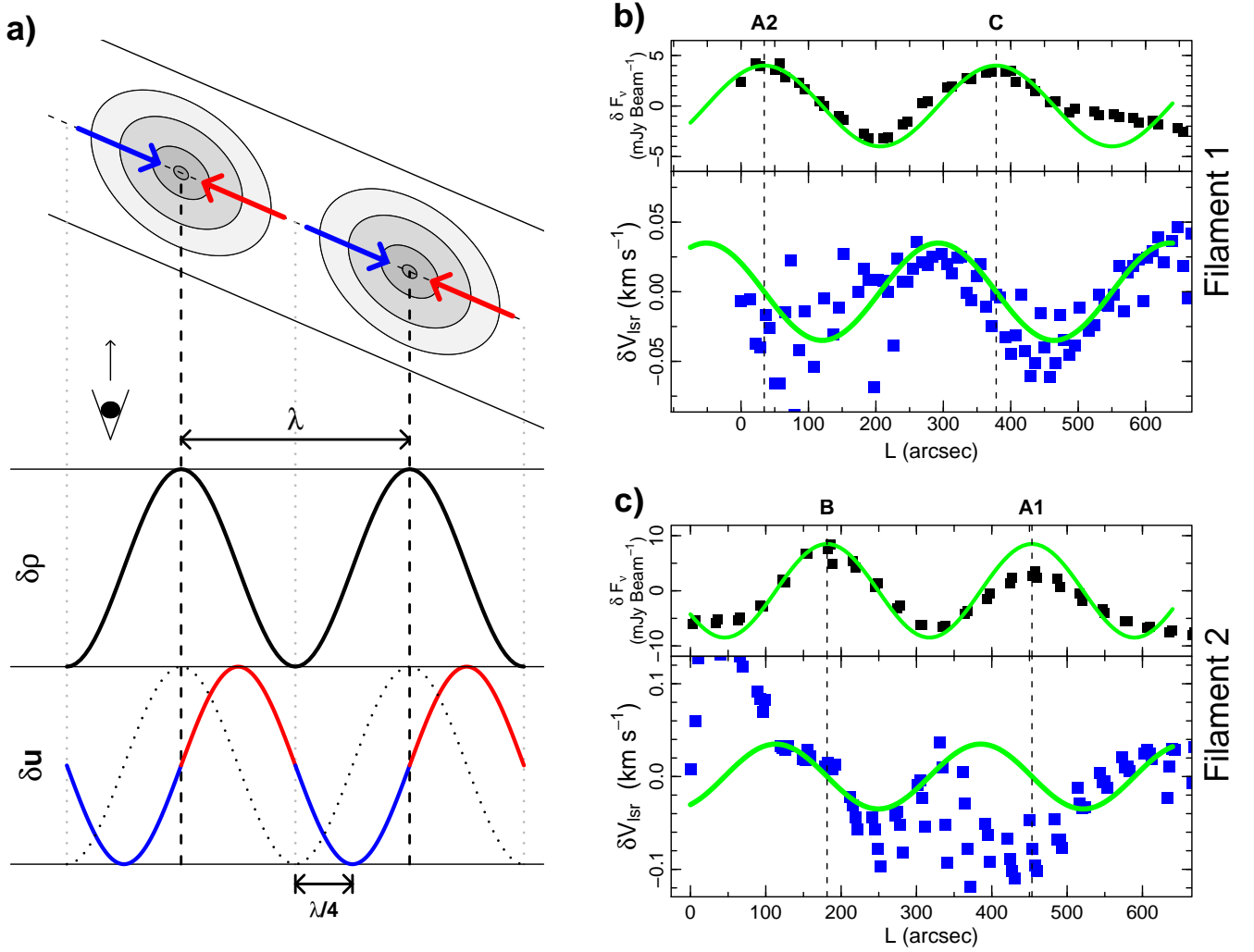}}
\caption{ Kinematic modeling of the velocity oscillations in filaments 1
and 2. {\bf a: } Schematic view showing how a core-forming
velocity field along the filament axis causes a
$\lambda$/4 shift between the sinusoidal perturbations of density and velocity. 
{\bf b } and {\bf c: } Comparison of the
density and velocity perturbations in filaments 1 and 2 
with the expectation from the simple kinematic model.
For each filament, the upper panel (black squares)
shows the increment of mm continuum flux over the mean (a proxy for
the density perturbation), and the lower panel (blue squares)
shows the variation in the C$^{18}$O centroid velocity over the
filament mean (after subtraction of a linear gradient). The
green solid line in each upper panel is a sinusoidal fit
to the density perturbation, from which a 
wavelength and phase
are determined. The green solid line in each bottom panel is
the result of shifting the density sinusoid by $\lambda$/4
(plus scaling it arbitrarily), and represents the expected 
pattern for a core-forming velocity field. 
The mean formal error in the $\delta V_{LSR}$ determination is 0.01~km~s$^{-1}$ for both filaments, 
which is on the order of the marker size.
\label{frag_model}}
\end{figure*}

Modeling in detail the fragmentation of the L1517 filaments
exceeds the scope of this paper, and would require 
information on so far unknown cloud properties 
like the strength and geometry of the magnetic field. 
For this reason, we limit ourselves to testing
whether the velocity oscillations of Fig.~\ref{FilVlsr} 
are consistent with
large-scale core-forming motions, and we do
so with a highly idealized infinite cylindrical 
model. Our first approximation is to 
only study motions along the
central axis of the cylinder, which allows us to convert the
problem to one dimension and which is motivated by
the dominance of longitudinal motions in simulations.
Following standard practice (e.g., \citealt{BIN87}, their 
section 5.1),
we use perturbation analysis and 
assume that the gas starts from a state of hydrostatic
equilibrium in which the 
density is constant along the filament axis (e.g., \citealt{OST64}).
To this equilibrium state we 
added a small longitudinal perturbation in both density and velocity, 
and work only to first order in the perturbation quantities (linear analysis).
Under these conditions, the equation
of continuity can be written as
\begin{equation}
\label{continuity}
{\frac{\partial\rho_1}{\partial t}} + \rho_0\; 
{\frac{\partial u_1}{\partial z}} = 0
\end{equation}
where $\rho$ is the density, $u$ the velocity, $z$ the
spatial coordinate along the filament axis, and the 0 and 1
subscripts refer to the unperturbed solution and the first-order
perturbation. If we now assume 
Fourier-component perturbations 
for both density and velocity, we can write 
\begin{equation}
\label{pert}
\rho_1(z,t) = C_1\; e^{i(kz -\omega t)} {\mathrm{~~~~~~and~~~~~~}} 
u_1(z,t) = C_2\; e^{i(kz -\omega t)},
\end{equation}
where $C_1$ and $C_2$ are constants due to the lack of
$z$-dependence of the unperturbed solution. Substituting these expressions
into the equation of continuity, we can obtain a simple relation between the
two constants.
The case of interest here is that of an unstable, core-forming mode
(i.e., opposed to a stable sonic wave), and this implies that
$\omega^2$ must be negative. Defining a real number $\gamma$ so
$\omega = i \gamma$, we can write
\begin{equation}
\label{phase_shift}
C_2 = i\; {\frac{C_1}{\rho_0}} \; {\frac{\gamma}{k}}.
\end{equation}
As $\rho_0$, $\gamma$, and $k$ are real numbers, the above equation 
implies that 
the Fourier component of the velocity perturbation must be 
shifted in phase from 
the density component by $\pi/2$, which is 
one quarter of the wavelength of the perturbation
(see also \citealt{GEH96}).
(In a stable (sonic) perturbation, $\omega$ is
a real quantity, so the density and velocity
Fourier components are in phase.)

The origin of the $\lambda$/4 shift between the 
velocity and density perturbations can be easily 
understood with the simplified cartoon 
of Fig.~\ref{frag_model}.a.
As can be seen, for the gas motions to
be core-forming, they have to 
converge towards the core centers, and this means that a
density peak must correspond to a position of vanishing velocity.
Assuming that both density and velocity perturbations are 
sinusoidal, this requires that there is 
a $\lambda$/4 shift between the two. 

The expected $\lambda$/4 shift between the density and the
velocity patterns in an
unstable perturbation provides a simple
criterion to test whether the
velocity field in the L1517 filaments is consistent 
with core-forming motions.
To this end, we have fitted a sinusoid to a cut of the mm-continuum emission 
along the filament axis. This quantity
should be proportional to the filament density profile, with the caveat
that the bolometer observation filters the extended emission due
to chopping, and from this fit we have determined the
wavelength
and phase of the density perturbation in filaments 1 and 2.
The top panels of Fig.~\ref{frag_model}.b and c show
that both filaments can be reasonably fitted
with sinusoidal density profiles that have wavelengths
of approximately 340 and 270~arcsec, 
respectively. 

To now test whether the 
observed velocity field
is consistent with the formation of these density profiles, 
we shift the density sinusoid by one quarter of the wavelength, 
as required by the perturbation analysis,
and compare the result with the velocity centroid data
towards the filament axis.
For this step, we have one degree of freedom in the choice of the
sign of the shift, as there are  two possible inclinations
of the filament with respect to the plane of the sky (e.g., the
eastern part of filament 1 could be inclined toward us or away from us),
and this adds a sign ambiguity to the gas radial velocity.
When this inclination is chosen (together with a scaling of
velocity amplitude), we derive the model predictions shown in the
bottom panels of Fig.~\ref{frag_model}.b and c 
superposed to the observed velocity centroids
(to which we have subtracted global linear gradients 
of 0.5~km~s$^{-1}$~pc$^{-1}$ in filament 1 and 
0.1~km~s$^{-1}$~pc$^{-1}$ in filament 2).

As can be seen, the velocity field of filament 1
agrees reasonably well with the prediction from 
the shifted sinusoid both in
wavelength and phase,
suggesting that the observed velocity oscillation is consistent
with core-forming motions.
The velocity towards core C does indeed coincide with a zero
value
in the velocity sinusoid, while the behavior of
the velocity towards core A2
is not as clear due to the higher scatter and the few
velocity points.
Such a reasonable fit is encouraging for an interpretation of
the velocity field in terms of fragmentation, as this filament
is also the best behaved 
according to both the maps (Fig.~\ref{C18O_channelmap}) and the radial
profiles (Fig.~\ref{Fils_RP}). It also harbors the chemically
youngest cores of the cloud (Fig.~\ref{cores_radial}),
suggesting that it is at the earliest phases of fragmentation.

In contrast to filament 1, filament 2
does not fit the shifted sinusoid
velocity pattern as well.
On the one hand, the velocity field of this filament 
does not follow a sinusoid pattern very closely, and on
the other, 
a region of high scatter between cores B and A1
leaves the velocity poorly defined, especially in 
the vicinity of core A1.
Still, as the figure shows, the position of 
core B lies in a region
consistent to have zero velocity,
which is the expected pattern for core-forming
motions.

So far we have concentrated our analysis on the phase of the
velocity oscillation, but the amplitude can also 
be measured and contains
information on the speed of the possible core-forming motions.
As illustrated by the cartoon in Fig.~\ref{frag_model}, 
the observed amplitude represents only the line-of-sight
component of the true velocity amplitude, and 
it therefore needs to be corrected for the
angle that the filament makes with the line of sight.
Lacking a better estimate, we assume
that each filament makes
a 45 degree angle with the line of sight, and we 
use this value to derive representative velocities of 
the possible core-forming motions.
As can be seen from Fig.~\ref{frag_model}, both filaments
have velocity oscillations with an amplitude of approximately
0.04~km~s$^{-1}$, which would correspond to a projection-corrected
value of approximately 0.06~km~s$^{-1}$.
Interestingly, such
subsonic speeds are similar to those inferred by
\citet{LEE99b} for the inward motions towards
a large number of low-mass dense cores, which were
obtained using a very 
different technique (analysis of self-absorbed profiles). 
Also in agreement with previous core lifetime
estimates \citep{LEE99a}, the time scale of collapse
inferred from the $\gamma$ value in Eq.~\ref{pert}
is on the order of 0.5~Myr.

Even if the velocity oscillations in L1517 do not arise from 
contraction motions, their amplitude
constrains core-formation models,
as any underlying contraction motion
has to be slower than about 0.05~km~s$^{-1}$
to remain undetected 
or confused with the observed oscillation. Thus, no matter how
we interpret the observed velocity pattern, 
the conclusion that core formation
in the L1517 cloud is strongly subsonic seems inescapable.

\section{How did the L1517 cores form?}\label{coreformation}

The picture that emerges from our analysis of
L1517 is that core formation in this cloud has involved only
subsonic motions and that it has started from conditions
that were very close to hydrostatic equilibrium
in an elongated configuration.
These characteristics, together with the 
close fit of the filament density profiles 
by the model of an 
isothermal pressure-supported cylinder, suggest that
some form of cylindrical gravitational fragmentation
has played a role in the formation
of the L1517 cores.

As discussed in section~\ref{simple_model}, the
simplest case of cylindrical fragmentation is that of an
isothermal, infinitely-long, pressure-supported cylinder.
Perturbation analysis shows that
the gravitational fragmentation of such a system has
a critical wavelength
$\lambda_f = 3.94 H$, where $H$ is the
filament width given by equation~\ref{h_diam}
\citep{STO63}.
Core formation via
this simplest gravitational fragmentation model
therefore requires that
the dense cores are physically spaced by
a distance grater than $\lambda_f$.

To test whether the separation between cores in L1517 is consistent
with the simplest gravitational fragmentation scenario,
we have estimated the critical wavelength 
of the filaments using the $H$ values obtained
from modeling their radial profiles (Table~\ref{Fils_fits}).
These values imply that $\lambda_f$ 
is about  $1.2 \times 10^{18}$~cm, 
or $570''$ for our assumed distance to Taurus (144~pc).
As can be seen from Fig.~\ref{frag_model}, the observed distance
between the cores in filaments 1 and 2 is  $\approx 300''$,
which is significantly less than the minimum value
expected from the theory.

Assuming that the filaments are inclined with respect to the
plane of the sky can change the above estimate, but 
it does not bring significant improvement.
On the one hand, correcting for projection will increase the true distance
between the cores, but it will also increase
$\lambda_f$ because the projection correction 
affects the column density
estimate from which $\lambda_f$ is derived (e.g., \citealt{ARZ11}).
As a result, the observed separation between cores in the L1517 filaments
seems to be about 1.5-2 smaller than
predicted by the simplest model of gravitational fragmentation.

A number of factors can potentially explain the smaller-than-predicted
separation between cores in L1517. First of all, our $\lambda_f$ 
value could have been overestimated. This is possible because $\lambda_f$
was derived from the density analysis of the filaments, which 
depends on assuming a standard C$^{18}$O abundance. If this
molecule has suffered from
large-scale freeze out in the filaments, the true gas volume density
would be higher, and $\lambda_f$ correspondingly smaller.
A second possibility is that edge effects, due to the finite 
size of the filaments, are important in the outcome of the fragmentation,
as suggested by the simulations of \citet{BAS83} and \citet{NEL93}.
Unfortunately, these effects depend very strongly on how sharp
the filament edges are \citep{NEL93}, and 
our data are not detailed enough to constrain this parameter.
Finally, magnetic fields are known to affect the fragmentation
of a cylinder
\citep{NAK93,NAK95,GEH96,HAN93,FIE00b}. \citet{FIE00b}, in particular,
show that both toroidal and poloidal  
magnetic fields 
decrease the critical length of fragmentation, and can do so
by a significant factor for sufficiently strong fields. 
(They also decrease the instability growth rate.) 
Whether this could explain the observations of L1517 is unclear, as
little is known about the magnetic field in the cloud.
\citet{KIR06} find that the field towards core B
is too weak to provide support,
but these measurements are limited
to a small region of the cloud already known to have undergone
core formation. A
detailed characterization of the large-scale magnetic field in the
L1517 cloud is clearly needed.

Even if the L1517 cloud is less symmetric than assumed by our model,
and its fragmentation history more complex than predicted by 
the simple theory, 
the observations presented here clearly constrain the process of
core formation to have involved an elongated
geometry and mostly subsonic motions. Core formation in
this environment therefore appears 
to be a two-step process in which the elongated configuration
close to equilibrium
is formed first and subsonic
fragmentation into cores occurs later.
This two-step scenario of core formation has already been proposed by
several authors based on different considerations,
and it therefore seems to apply to more regions than just
the one studied here \citep{SCH79,HAR02,MYE09}.
What our L1517 data show now is that
the filamentary gas prior to fragmentation has a
subsonic level of turbulence and a coherent velocity
pattern, two properties that were thought before to
only apply to dense cores. For
such a quiescent state of the gas
to occur, turbulent motions must
have dissipated prior to (or rapidly during) filament
formation, and therefore must play a limited
role in the formation of the individual cores.
If such a scenario is representative
of core formation in other dark clouds,
subsonic, velocity-coherent filaments must be
a common feature, and embedded cores must systematically present
kinematic patterns continuously connected to those
of the large-scale filaments.
Detailed study of the connection between dense
cores and their surrounding cloud
in a larger sample of regions
should be able to test these two predictions.

\section{Summary}

We observed the L1517 dark cloud in C$^{18}$O(1--0), N$_2$H$^+$(1--0),
SO(J$_{\mathrm{N}}$=3$_2$--2$_1$) with the
FCRAO telescope, and in the 1.2mm dust continuum with the IRAM 30m 
telescope. From the analysis of these data, we came to the 
following main conclusions.

1. The gas in the L1517 cloud is structured in four filaments with
typical sizes of about 0.5~pc.

2. The radial profile of C$^{18}$O(1--0) emission in each filament
consists of a central flattened region and a power-law tail. An
isothermal cylinder model provides an approximate fit to the emission,
although better agreement is obtained with a softened power law profile.

3. Five starless cores are embedded in the filaments. Their chemical
composition indicates that they are at different evolutionary stages,
although their kinematic properties are very similar.

4. The velocity field of the gas in the L1517 cloud has been characterized by
fitting Gaussians to the observed spectra. Most positions require one
Gaussian, although a few regions require two, very
likely due to the overlap between different components.

5. The filaments are extremely quiescent. Their nonthermal 
linewidth is subsonic and changes very little over the length
of the filaments. The velocity centroids also change subsonically
over the filament length.
These characteristics 
indicate that the gas in the filaments is velocity coherent
on scales of about 0.5~pc.

6. Although quiescent, the filaments have large-scale patterns of
(subsonic) velocity. The gas in the dense cores follows
the velocity pattern of the less dense gas  closely, indicating that core
formation has not decoupled the dense
gas kinematically from its surrounding filament material.

7. In two filaments, the large-scale velocity patterns seem
to consist of  oscillations. A simple kinematic model shows that, 
at least in one filament, the oscillatory pattern is consistent
with core-forming motions along the axis of the filament.

8. Core formation in L1517 seems to have occurred in two steps.
First, the subsonic, velocity-coherent filaments have condensed
out of the more turbulent ambient cloud. Then, the cores have
fragmented quasi-statically and inherited the kinematics
of their parental filament. Turbulence dissipation has therefore occurred
on scales of 0.5~pc or larger, and seems to have played
little role in the formation of the individual cores.

\begin{acknowledgements}
We thank Mark Heyer for assistance with the FCRAO observations, 
Gilles Duvert for providing us with his old CO observations
of the L1517 region, and
Jouni Kainulainen for communicating his unpublished 
extinction map of L1517.
We also thank an anonymous referee, Malcolm Walmsley, and Jens Kauffmann 
for a number of 
comments and suggestions that helped clarify the presentation.
This research made use of NASA's Astrophysics Data System 
Bibliographic Services and the SIMBAD database,
operated at the CDS, Strasbourg, France.
It also made use of EURO-VO software, tools, and services. 
The EURO-VO is funded by the European Commission through 
contracts RI031675 (DCA) and 011892 (VO-TECH) under the 6th Framework 
Program and contracts 212104 (AIDA) and 261541 (VO-ICE) under 
the 7th Framework Program.
The Digitized Sky Survey was produced at the Space Telescope Science 
Institute under U.S. Government grant NAG W-2166. The images of these 
surveys are based on photographic data obtained using the Oschin Schmidt 
Telescope on Palomar Mountain and the UK Schmidt Telescope. The plates 
were processed into the present compressed digital form with the 
permission of these institutions. 
\end{acknowledgements}

\end{document}